\PassOptionsToPackage{dvipsnames}{xcolor}

\documentclass[acmsmall,screen]{acmart}

\setcopyright{none}
\copyrightyear{\textcopyright{} Zeyu Huang, Xinyi Cao, Yue Deng, Junze Li, Kangyu Yuan, Xiaojuan Ma, 2025. This preprint is for your personal use and not for redistribution. The definitive Version of Record was published in ACM Digital Library at \url{https://doi.org/10.1145/3799444}}
\acmJournal{PACMHCI}
\acmYear{2026} \acmVolume{10} \acmNumber{1} \acmArticle{GROUP009}
\acmMonth{3} \acmDOI{10.1145/3799444}

\acmSubmissionID{2923} 

\received{July 2025}
\received[revised]{November 2025}
\received[accepted]{December 2025}

\usepackage{rotating} 
\usepackage{array}
\usepackage{makecell}
\usepackage[inline]{enumitem}
\usepackage{xspace}
\usepackage{multirow} 
\usepackage{subcaption} 
\usepackage{hyperref}

\newcommand{\name}{Drop'n'Collect\xspace}
\newcommand{\edit}[1]{{#1}\xspace}

\newcommand{\atconnect}{@connect.ust.hk}

\begin{document}
\title[Grassroots Understanding and Practices of Collective Memory Co-Contribution]{Exploring the Grassroots Understanding and Practices of Collective Memory Co-Contribution in a University Community}

\author{Zeyu Huang}
\email{zhuangbi\atconnect}
\orcid{0000-0001-8199-071X}
\affiliation{%
  \institution{The Hong Kong University of Science and Technology}%
  \state{Hong Kong}%
  \country{China}%
}%

\author{Xinyi Cao}
\email{essiecao@stanford.edu}
\orcid{0009-0005-8601-5039}
\affiliation{%
  \institution{Stanford University}%
  \streetaddress{475 Via Ortega}%
  \city{Stanford}%
  \state{California}%
  \country{USA}%
  \postcode{94305}%
}

\author{Yue Deng}
\email{ydengbi\atconnect}
\orcid{0009-0008-5339-8792}
\affiliation{%
  \institution{The Hong Kong University of Science and Technology}%
  \state{Hong Kong}%
  \country{China}%
}%

\author{Junze Li}
\email{jlijj\atconnect}
\orcid{0000-0003-4768-6495}
\affiliation{%
  \institution{The Hong Kong University of Science and Technology}%
  \state{Hong Kong}%
  \country{China}%
}%

\author{Kangyu Yuan}
\email{kyuanaf\atconnect}
\orcid{0009-0001-8460-9651}
\affiliation{%
  \institution{The Hong Kong University of Science and Technology}%
  \state{Hong Kong}%
  \country{China}%
}%

\author{Xiaojuan Ma}
\authornote{Corresponding author}
\email{mxj@cse.ust.hk}
\orcid{0000-0002-9847-7784}
\affiliation{%
  \institution{The Hong Kong University of Science and Technology}%
  \state{Hong Kong}%
  \country{China}%
}%

\renewcommand{\shortauthors}{Zeyu Huang et al.}

\begin{abstract}
  Collective memory---community members' interconnected memories and impressions of the group---is essential to the community's culture and identity. Its development requires members' continuous participatory contribution and sensemaking. However, existing works mainly adopt a holistic sociological perspective to analyze well-developed collective memory, less focusing on member-level conceptualization of this possession or what the co-contribution practices can be. Therefore, this work alternatively adopts the latter perspective and probes such interpretative and interactional patterns with two mobile systems. With one being a locative narrative and exploration system condensed from existing literature's design frameworks, and the other being a conventional online forum representing current practices, they served as the anchors of observation for our two-week, mixed-methods field study (n=38) on a university campus. A core debate we have identified was to retrospectively contemplate or document the presence as a history for the future. This also subsequently impacted the narrative focuses, expectations of collective memory constituents, and the ways participants seek inspiration from the group. We further extracted design considerations that could better embrace the diverse conceptualizations of collective memory and bond different community members together. Lastly, revisiting and reflecting on our design, we provided extra insights on designing devoted locative narrative experiences for community-driven UGC platforms.
\end{abstract}



\begin{teaserfigure}
  \centering
  \begin{subcaptionblock}[c]{0.3\textwidth}
    \centering
    \includegraphics[height=8cm]{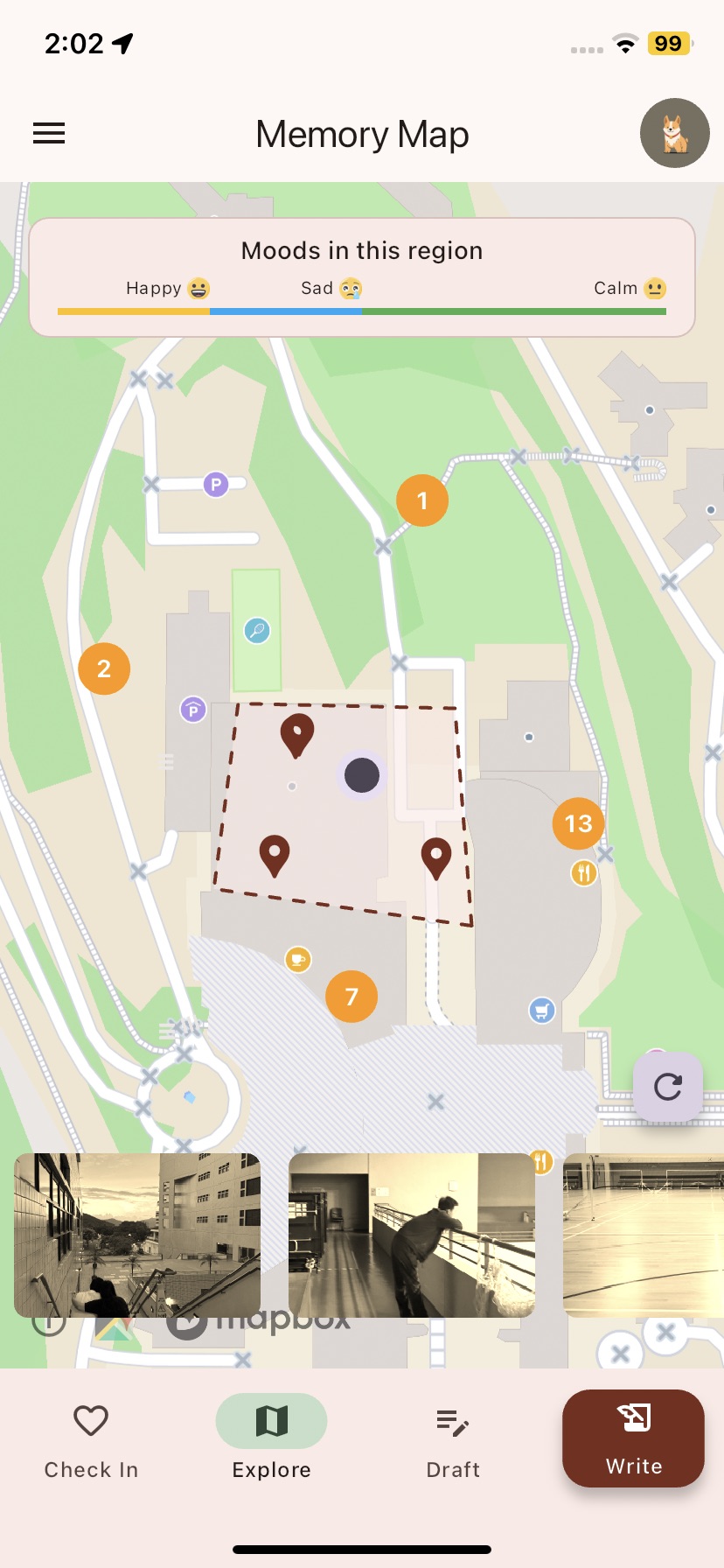}
    \caption{Exploring memories on a map}
  \end{subcaptionblock}
  \begin{subcaptionblock}[c]{0.68\textwidth}
    \centering
    \includegraphics[width=0.49\textwidth]{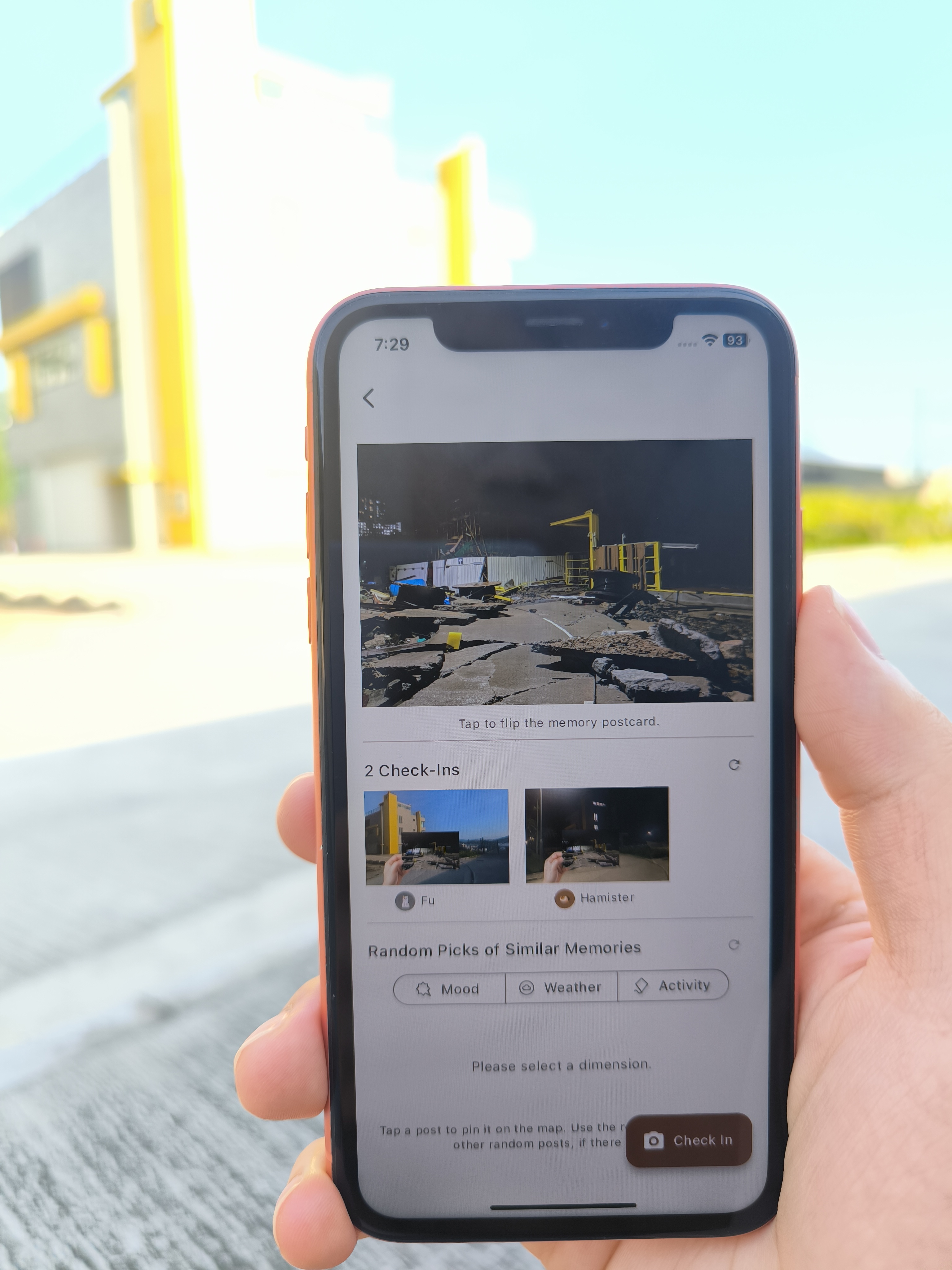}
    \includegraphics[height=8cm]{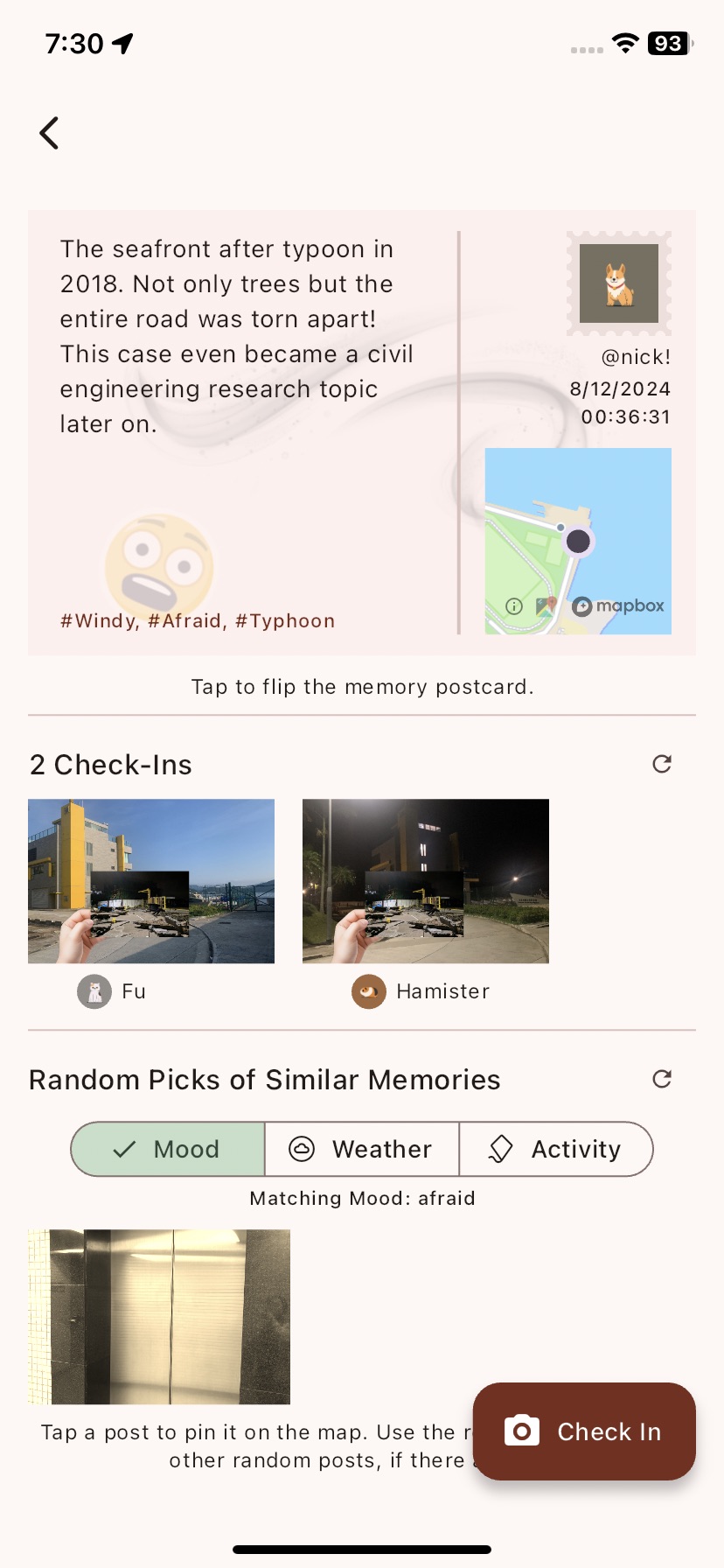}
    \caption{A real-world use scenario}\label{fig:real-world}
  \end{subcaptionblock}
  \caption{\name, one of our design probes to explore HCI support for collective memory co-contribution among co-located university members. It inherits locative narrative design for memory sharing, contains metaphorical elements for memory postcards, and advocates serendipitous memory discovery.}
  \Description{(a) The screenshot of \name home page. A map view shows the current location at the center and three location pins to indicate nearby posts. A horizontal scrolling list displays three memory postcards in sepia filter. Some orange circled numbers on the map show how many posts are in adjacent areas. (b) Two photos. The first is a user holding a mobile phone in front of a yellow building on the road. The mobile phone shows a memory postcard, which is exactly the same spot but the building and the road are destroyed by natural disasters. The second is a screenshot of the system showing the same memory post. It shows the flip side of the memory postcard, which is a user's memory about a typhoon.}
  \label{fig:teaser}
\end{teaserfigure}


\maketitle

\section{Introduction}\label{sec:intro}


Collective memory---originally a sociological construct---describes the active co-contribution, exchange, and visiting of community members' individual memories~\cite{twymanBlackLivesMatter2017,harrisCollaborativeRecallCollective2008,wegnerTransactiveMemoryContemporary1987,jonesDesigningSocialMemory2016}
that are related to the shared experiences of the community~\cite{sasDesigningCollectiveRemembering2006,ijabsCollectiveMemory2014}.
As an essential intangible possession of a community~\cite{harrisCollaborativeRecallCollective2008}, collective memory differs from a simple collection of personal memories because it also emphasizes the interconnections among members' shared experiences and life episodes, as well as how the interconnections benefit the community as a whole~\cite{sasDesigningCollectiveRemembering2006, harrisCollaborativeRecallCollective2008}.
On the one hand, it provides an ever-updating lens into the present conception of a group's past, cultures, and vibes~\cite{roedigerCollectiveMemoryNew2015}.
On the other hand, it stimulates members to reflect, prospect, bond with each other~\cite{sasDesigningCollectiveRemembering2006}, and eventually form group identities~\cite{harrisCollaborativeRecallCollective2008,eyermanPresentCultureTransmission2004,orianneCollectiveMemoryIndividual2023}.


In the HCI and CSCW fields, we observe that most current work focuses on sociological analysis of existing communities' collective memory~\cite{twymanBlackLivesMatter2017,ferronCollectiveMemoryBuilding2011,jonesCoconstructingFamilyMemory2018}.
They primarily see collective memory as a well-developed product through a holistic view of the entire community---what we refer to as an outcome-oriented, ``top-down'' perspective.
What we think can generate complementary HCI insights is a process-oriented, ``bottom-up'' perspective of inquiry that prioritizes grassroots conceptualizations. For example, such inquiry can include community members' own understanding of what constitutes collective memory, how collective memory inspires them, and how members practically co-contribute to collective memory in the everyday context.
They shall provide more human-centered design insights for facilitating the community's collective memory development rather than experts' understandings.
Therefore, this paper aims to practice this ``bottom-up'' perspective and uncover grassroots understanding and practices of this concept.

We achieve this by explicitly engaging them in collective memory co-contribution with research probes to facilitate the process.
During their conscious exposure to this concept and the digitally supported interactions, we inquire into the emerging behavior and interpretative patterns, as well as the rationales behind them.


\edit{
  In designing our research probes, we observed that the existing literature highlights collective memory development across two system modalities.
  One is the well-established content-oriented design of social platforms, such as online forums~\cite{glawionRememberingWorldWar2023}.
  The other, while not so widely applied in the real world, emphasizes incorporating newer technologies, engaging memory-related interactions, and foregrounding interconnections among memories~\cite{velhinhoDesigningCollaborativeStorytelling2024,felasariCollatingCitysCollective2021}.
  Therefore, we adopted a dual-system study setup to investigate whether different system designs influence people's content generation practices and collective memory interpretations, and to ensure generalizability by accounting for potential preference variation.
  We designed a conventional content-oriented online forum system, as well as another locative and memory-skeuomorphic system derived from recent literature's frameworks~\cite{liuGrassrootsHeritageCrisis2010,lakCollectiveMemoryUrban2019}.
  And we deployed both systems as a between-subject study.
}
Despite the dual system setup, we did not aim to find a better design from the two.
\edit{Instead, we took an exploratory approach, using these probes to uncover people's conceptualization of and interactions surrounding collective memory, as well as how different features might or might not support its development.}
\edit{Specifically, we are guided by the following research questions:
  \begin{enumerate}[label=\textbf{RQ\arabic*}]
    \item When prompted to contribute to their community's collective memory, how do members conceptualize this construct, and what content do they deem relevant?\label{rq:conceptualization}
    \item When practicing collective memory co-contribution, how may technological probes introduce variances in people's content generation behaviors and interpretations of the memory output?\label{rq:practice}
    \item How can the design of future systems better support and bridge the diverse interpretations and practices of grassroots collective memory co-contribution?\label{rq:considerations}
  \end{enumerate}}

We deployed both systems at a local university. Thirty-eight university members were invited to a field study, assigned either system, and explicitly asked to collaboratively contribute to the collective memory of the campus community with the assigned system. And our evaluation contained observations, exit-interviews, and questionnaires, our study.

\edit{As a result, our study yields three major contributions.
  \begin{enumerate}[label=\arabic*.]
    \item We identify key conceptual tensions in grassroots collective memory co-contribution---e.g., retrospective contemplation versus documenting the present, informative knowledge versus relatable thoughts.
    \item We present diverse user practices of memory generation, as well as their interpretation of memory outputs, that future designs should support.
    \item We offer several design considerations for future systems intended to foster community-driven memory exchange.
  \end{enumerate}
}
At a higher level, we encourage HCI designers to acknowledge and actively embrace these distinctions in interactional and interpretative patterns, and strive to bond different community members together through inclusive designs.


\section{Related Work}

\subsection{Collective Memory}\label{sec:rw:cm}
Collective memory is originally a sociological construct describing shared conceptions of a community's past and group members' relations to it~\cite{heuxCollectiveMemoryAutobiographical2023}.
Although the precise definitions can vary with the disciplines and domains of analysis~\cite{harrisCollaborativeRecallCollective2008,heuxCollectiveMemoryAutobiographical2023}, two essential elements in consensus are the interconnections among individual memories about the community and the dynamics of people interacting with each other's memories to form a shared overall understanding of the community eventually~\cite{wegnerTransactiveMemoryContemporary1987}.
The first element---the interconnections---can be reflected by spatial or temporal similarities of memories~\cite{halbwachsSpaceCollectiveMemory1950},
or that the memories are related to similar occurrences~\cite{sasDesigningCollectiveRemembering2006}.
For example, the collective memory of people in the same neighborhood is anchored by their similar experiences in the same physical surroundings, such as worship at the same church, trades at the same marketplace, etc.~\cite{halbwachsSpaceCollectiveMemory1950}.
Such shared context connects different personal narratives together,
resulting in a combined micro- and macro-level overview of the community instead of a mere collection of narrative pieces~\cite{wegnerTransactiveMemoryContemporary1987}.
The second element---the dynamics---stresses the massive amount of visiting, exchanging, and complementing the shared knowledgebase, during which community members gain a deeper understanding of each other and the community as a whole~\cite{twymanBlackLivesMatter2017,harrisCollaborativeRecallCollective2008,wegnerTransactiveMemoryContemporary1987,jonesDesigningSocialMemory2016}.

As briefed in \hyperref[sec:intro]{Introduction}, we categorize collective memory research into ``top-down'' and ``bottom-up'' perspectives.
The former, being the mainstream adoption by sociologists, psychologists, and historians, holistically and retrospectively anatomizes a community with already-formed collective memory, often observing the group's eventual remembrance in an outcome-oriented manner (e.g.~\cite{schwartzAbrahamLincolnForge2003,twymanBlackLivesMatter2017,halbwachsSpaceCollectiveMemory1950,wegnerTransactiveMemoryContemporary1987,ricoeurPersonalMemoryCollective2006}).
Alternatively, the ``bottom-up'' and process-oriented perspective focuses on the grassroots practices of collective remembrance and understanding of collective memory as a shared possession~\cite{hirstCollectiveMemoryPsychological2018,heuxCollectiveMemoryAutobiographical2023}.
This perspective is less adopted in existing studies~\cite{heuxCollectiveMemoryAutobiographical2023}.
However, we would like to underline its potential for uncovering more first-person perceptions and opinions on collective memory co-construction, as well as echoing the emerging appeals for more human-centered designs that practically support the development of collective memory~\cite{jonesDesigningSocialMemory2016,vanhouseTechnologiesMemoryKey2008,liuGrassrootsHeritageCrisis2010}.

\subsection{HCI Approaches to Collective Memory}\label{sec:rw:hci-cm}

Researchers and practitioners seek to observe and facilitate the curation of collective memories, given their importance to community-building.
In the HCI field, there seems to be a disparity in the target scenarios studied by social computing scholars and system designers.

The former tends to stay in sync with sociologists' and psychologists' emphasis on the top-down investigation into existing collective memories~\cite{harrisCollaborativeRecallCollective2008,wegnerTransactiveMemoryContemporary1987,sasDesigningCollectiveRemembering2006,ijabsCollectiveMemory2014}.
For example,
Ferron et al.\@ and Twyman et al.\@ conducted two studies
on how communities collectively remember critical social movement events on Wikipedia~\cite{twymanBlackLivesMatter2017, ferronCollectiveMemoryBuilding2011}.
Two other studies investigated how collective memory was formed in families~\cite{jonesCoconstructingFamilyMemory2018} and trauma teams~\cite{sarcevicTransactiveMemoryTrauma2008}.
They have highlighted the importance of digitalizing and preserving personal narratives and fostered the interactions among community members, but have discussed less how to practically tackle the fragmented nature of memory pieces.
Also, they focused less on communities' daily routines and ordinary experiences, another organic part of the collective memory~\cite{halbwachsSpaceCollectiveMemory1950}.

In comparison, Jones et al.\@ propose to approach collective memory as a procedure of continuous ``(re)construction and (re)interpretation'', especially the use of artifacts to facilitate ``creating, sharing, and revisiting'' of the collective memory~\cite{jonesDesigningSocialMemory2016}.
Accordingly, relevant designs strove to enable effective memory aggregation and retrieval within communities~\cite{hakkilaConnectingEvaluatingIndigenous2023,klaebeDigitalStorytellingHistory2007,lindleyNarrativeMemoryPractice2009}, as well as integrated memory authoring and exploration flow~\cite{wegnerTransactiveMemoryContemporary1987}.
Uriu et al.\@ fostered serendipitous discoveries of memories in the extended family tree~\cite{uriuCaraClockInteractivePhoto2009}. Although it targeted small-scale kinship interactions, it has demonstrated the value of serendipity and assisted roaming over the collectivity.
Cheverst et al.\@ showed the advantages of locative media experience for collective memory with a locative authoring and exploration application for a rural village community~\cite{cheverstSupportingConsumptionCoauthoring2017}. Being situated in the physical surroundings created a more immersive and empathetic experience of collective memory exploration~\cite{cheverstSupportingConsumptionCoauthoring2017}. Nonetheless, their work advocated an expert-directed curation of collective memory, and their evaluation mostly covers usability topics~\cite{cheverstSupportingConsumptionCoauthoring2017}.
Another system by Cranshaw et al.\@ persuading travelers to journal for less prominent places in a travel destination~\cite{cranshawJourneysNotesDesigning2016}. Though they provided exemplary designs of an interactive and intentionally poetic system to engage the community, their evaluation did not cover how the design actually influenced the journaling experience.

In sum, community members' own perception of collective memory is less discussed in current works, nor its relationships with the community's interactions~\cite{halbwachsSpaceCollectiveMemory1950}.
This is exactly the focus of this paper, and we hope to ultimately reinforce the bridge between the theoretical framing of collective memory and the grassroots efforts in its active development.


\subsection{Individual Narratives of the Past}

There are other HCI works that support the recording and revisiting of individual memories, using forms such as tiles~\cite{bennettTopoTilesStorytellingCare2015}, buttons~\cite{riggsDesigningArchiveFeelings2024}, and paintings~\cite{wanMetamorpheusInteractiveAffective2024} as symbolic representations and proxies of personal memories, and digital replications of past experiences in AR~\cite{liExploringOpportunitiesAR2023,merrimanFamiliarEnvironmentsEnhance2016}.
Their main focus was the outcome itself as individual memorial artifacts that are circulated within a restricted scope. They did not emphasize the dynamic, transactive process of a community's memories and the formation of an integrated conception.
Still, these studies provided implications on how to stimulate memory with situated environments~\cite{bennettTopoTilesStorytellingCare2015}, photos~\cite{liExploringOpportunitiesAR2023}, and memory-related objects~\cite{riggsDesigningArchiveFeelings2024}.

Meanwhile, we note that social media sites are also common places to publish and archive personal experiences, as well as to disseminate them within the shared social network.
However, they are still designed to highlight people's individual lives at the moment~\cite{louieOpportunisticCollectiveExperiences2021} with limited advocacy for memory contents that are meaningful to a community or for interpreting contents from the community's perspective~\cite{naamanItReallyMe2010}.

Although these designs focus less on the communal and social aspects of collective memory, they provide valuable design examples and perspectives about memories for our work.

\section{Prototyping Mobile Apps for Collective Memory Co-Contribution}\label{sec:design}

As mentioned in \autoref{sec:intro}, considering the transactive and social nature of collective memory~\cite{wegnerTransactiveMemoryContemporary1987}, we employed a dual-system setup to both respect common existing practices and explore recent design proposals and frameworks.
We adopted both a conventional forum-like system and a dedicatedly designed system (namely \name{}), in the hope of better probing into users' interpretations of collective memory, practices relevant to its co-contribution, and design opportunities to facilitate the process. 
The next two subsections cover our design process for \name. While the forum system did not receive a dedicated design process (because it aligns with the common forum user experience), we ensured that both systems' features paired up to serve the same set of user intents. The final features of both systems are described pairwise in the last subsection.

\subsection{Design Goals and Requirements for \name}\label{sec:design-goals}

Referring to Liu's framework~\cite{liuGrassrootsHeritageCrisis2010}, we highlighted the following five design goals for HCI systems to support ``socially-distributed curation'' of collective memories:
\begin{enumerate*}
\item active participation at the individual level,
\item content collection at the community level,
\item continuous active efforts across time,
\item meaning-making of the interconnected memories, and
\item interactive memory sharing.
\end{enumerate*}

\begin{figure}[ht]
  \centering
  \includegraphics[width=\textwidth]{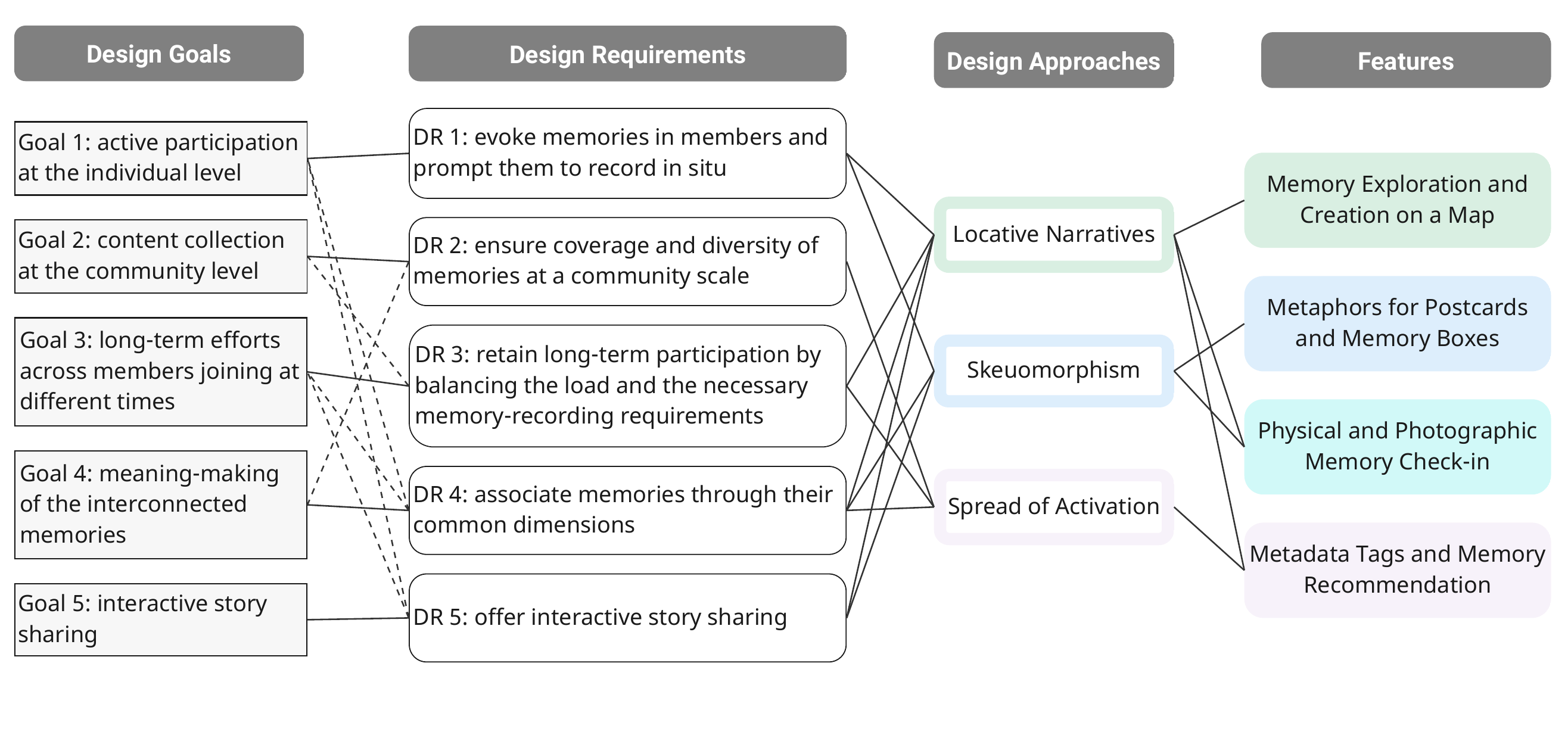}
  \Description{A diagram with four columns: "Design Goals", "Design Requirements", "Design Approaches", and "Features". The items and the linkage between them are stated in the body text.}
  \caption{An overview of the design process for \name.}
  \label{fig:design-process}
\end{figure}

We first translated these high-level goals into actionable design requirements (\autoref{fig:design-process}).
To start with, community members' active contribution (Goal 1) is intrinsically motivated upon recollecting their community-related experiences and feelings~\cite{cranshawJourneysNotesDesigning2016}. As the recollected episodes and emotions may disappear when interrupted~\cite{viskontasNeuralCorrelatesRecollection2009}, it is important to nudge people to put them down when freshly in the mood. We thus summarized the design requirement: \textbf{DR1.\@ evoke memories in members and prompt recording in-situ}.
Second, a more complete picture of the group can be obtained by involving more people to contribute (Goal 2), as every member's memory only reflects certain facets of the community~\cite{wegnerTransactiveMemoryContemporary1987, heuxCollectiveMemoryAutobiographical2023}. In this regard, we proposed: \textbf{DR2.\@ ensure coverage and diversity of memories at a community scale}.
Third, a living collective memory requires continuous contributions of active members (Goal 3).
If the task is very demanding or boring, it is difficult to maintain prolonged involvement. This led to \textbf{DR3.\@ retain participation by balancing the load and the necessary memory-recording requirements}.
Fourth, to ensure that individual memories are interconnected in the community, it is critical to encourage community members to seek meanings from such connections and consciously select individual memories (Goal 4). The most intuitive way to achieve it is by situating an individual's memory in the community's unique context. We framed it as \textbf{DR4.\@ associating memories through their common dimensions (e.g., locations, activities)}.
Fifth, from the user experience perspective, designing an interactive story-sharing experience (Goal 5) could enhance listeners' engagement and emotional impact~\cite{pittarelloDesigningContextAwareArchitecture2011}.
We kept it straightforward as \textbf{DR5.\@ interactive story sharing}.
A design requirement can also meet multiple goals. For example, ensuring memory coverage (DR2) may lay a foundation for meaning-making of memories' interconnections (Goal 4); easing the system's load (DR3) may invite more concurrent participation as well (Goal 2); connecting members memories (DR4) may help members see the value of their contributions and consequently encourage their participation (Goal 1) and retention (Goal 3). Interactive sharing (DR5) can become engaging enough to motivate individuals' and the community's maintained participation (Goals 1 \& 3).

\subsection{Design Approaches for \name}\label{sec:design:approach}

Next, we identified three approaches to circumscribe the features of a collective memory support system to meet the aforementioned design requirements. They are: locative narratives, skeuomorphism, and spread of activation.

Locative interactions, such as Pok\'emon Go\footnote{\url{https://pokemongo.com/}, a mobile game}, pin interactive elements to specific geolocations and require users to physically go there. And locative narratives are those leveraging locative interactions for content consumption and creation~\cite{nisiAuthoringLocativeNarratives2022}.
As mentioned in \autoref{sec:rw:hci-cm}, they can offer situated content consumption and an engaging and entertaining user experience through physically engaged interactions.
It is also adopted by commercial platforms such as AllTrails\footnote{\url{https://www.alltrails.com/}, a hiking experience sharing and trail authoring platform.}, as well as HCI works for history preservation~\cite{cheverstSupportingConsumptionCoauthoring2017}, interactive fiction~\cite{millardTireeTalesCooperative2017,millardCanyonsDeltasPlains2013}, random short messages~\cite{liuPinsightNovelWay2018,cranshawJourneysNotesDesigning2016,juBreadcrumbSNSAsynchronous2015}, and specialized information~\cite{changTourgetherExploringTourists2020,siriarayaPalmHappinessLocationBased2022}.
Thus, we anticipate that locative narratives achieve a similar effect of meaningful content sharing in the context of collective memory co-contribution (\textbf{DR5}).
In addition, memories are often strongly related to specific locations~\cite{bennettTopoTilesStorytellingCare2015,tuanTopophiliaStudyEnvironmental1990}.
Being situated in the location can trigger people's recalling of memories and the accompanying emotions~\cite{tuanTopophiliaStudyEnvironmental1990} (\textbf{DR1}).
Lastly, location information is an essential feature of collective memory that holds ``environments'' and ``meanings'' beyond mere space~\cite{lakCollectiveMemoryUrban2019}.
It can serve as a link across spatially related memory pieces and the bearer of people's feelings and thoughts~\cite{tuanTopophiliaStudyEnvironmental1990} (\textbf{DR4}).

Skeuomorphism refers to the UI design technique of imitating similar artifacts in another material~\cite{pageSkeuomorphismFlatDesign2014}.
It is frequently used in HCI designs for personal reflection~\cite{sunPostcardYourFood2020,gerritsenMailingArchivedEmails2016}, memory anchors~\cite{shojiMuseumExperienceSouvenir2024,kangTieMemoriesEsouvenirs2022}, and so on.
For example, the postcard is an effective skeuomorphic object to facilitate the reflection of the past~\cite{sunPostcardYourFood2020,gerritsenMailingArchivedEmails2016}.
Encompassing the abstract concept of memory in familiar and everyday skeuomorphic concepts not only allows engaging interactions (\textbf{DR5}) but also stresses the meaningfulness of their sharing (\textbf{DR1}).


Finally, we borrowed the construct \emph{``spread of activation''} from psychology to describe the mechanism of inciting and discovering more associated memory content within the community. This was intended to complement locative narratives that center primarily on memories about a particular spatial region.
As a cue to direct the users (i.e., how the ``network'' is connected in the analogy,  \textbf{DR4}), we leveraged Lak et al.'e framework of collective memory's core elements, such as locations' meanings, things people do, personal contexts, and history over time, as they are the natural cues that associate individual memories in a community~\cite{lakCollectiveMemoryUrban2019}.
Analogous to ``activating'' a selection of ``source nodes'' and propagating the ``activation'' to other ``nodes in a network''~\cite{crestaniSpreadActivationTheory2012}, we wanted to engage them in exploring other related memories that were spatially and temporally distributed, and raise their awareness of the multi-facet nature of collective memory. Consequently, it could enhance the coverage and diversity of their conceptualization of the collective memory and their contribution to it (\textbf{DR2}).

After pinpointing these high-level approaches, we specifically brainstormed skeuomorphic elements because this design approach did not lead to concrete interactive features as opposed to the other two. We aimed to find elements that, while delivering metaphors, were also in harmony with other features. For example, a candidate feature was to visualize memory posts as Mixed-Reality (MR) flowers on the ground, which users can plant or pick up. This candidate was not selected due to the substantial outdoor inaccuracy of locative MR and the time-consuming prerequisite of scanning planes. Another exemplar candidate was MR mailboxes for lifelike deposit and withdrawal of memory postcards. While rejected for similar reasons, it inspired our final, more lightweight alternative, which is described below.

\subsection{Features and Implementation}\label{sec:design:features}

\begin{figure*}
  \hfill
  \begin{subcaptionblock}[t]{0.4\textwidth}
    \centering
    \includegraphics[width=0.83\textwidth]{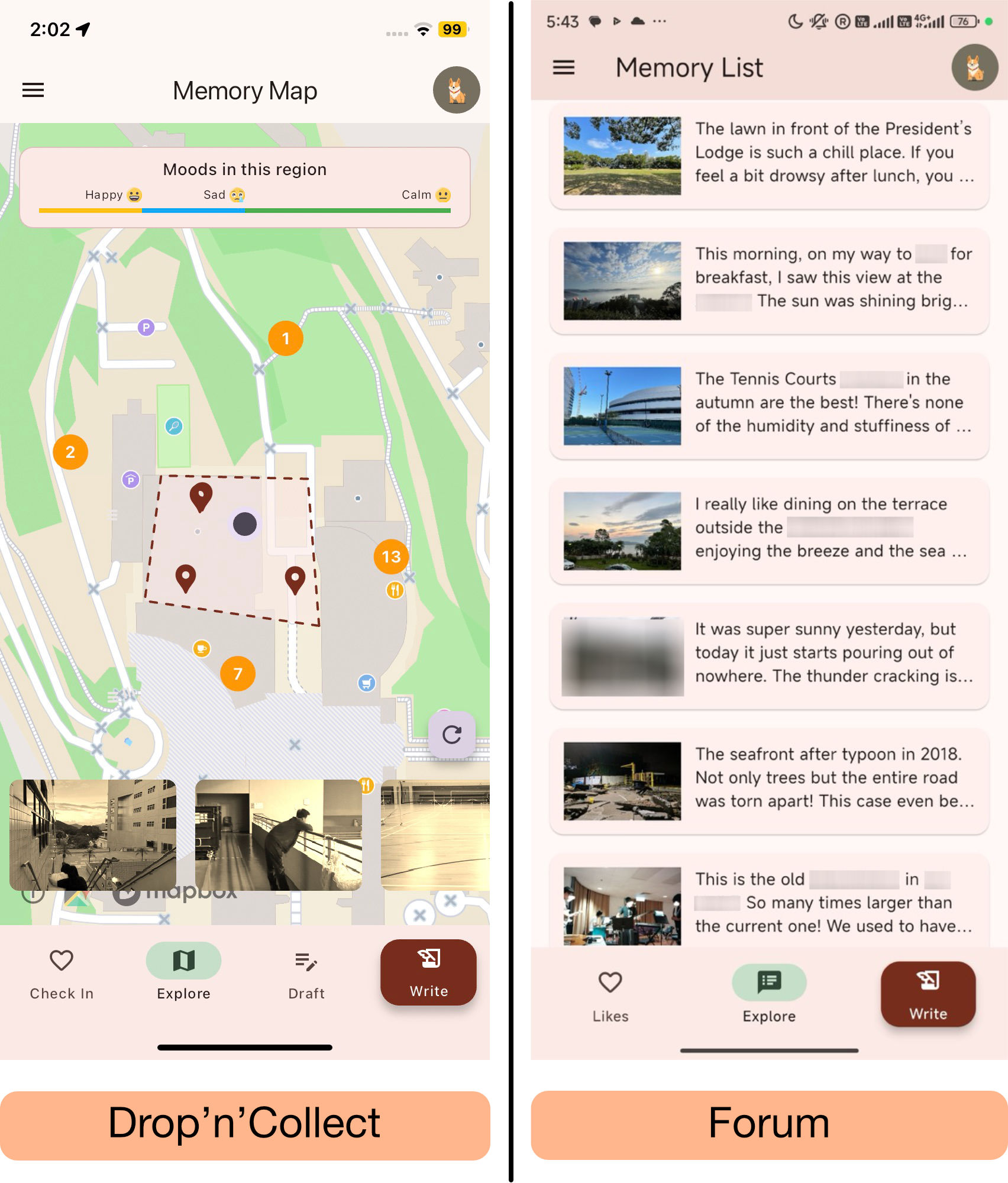}
    \caption{Browsing memories in both systems}\label{fig:browse}
    \Description{The left screenshot shows the map view of \name with pins of nearby posts' locations and a horizontal list of their photos. The right screenshot shows the list view of the forum with every post's photo and content preview.}
  \end{subcaptionblock}
  \hfill
  \begin{subcaptionblock}[t]{0.5\textwidth}
    \centering
    \includegraphics[width=\textwidth]{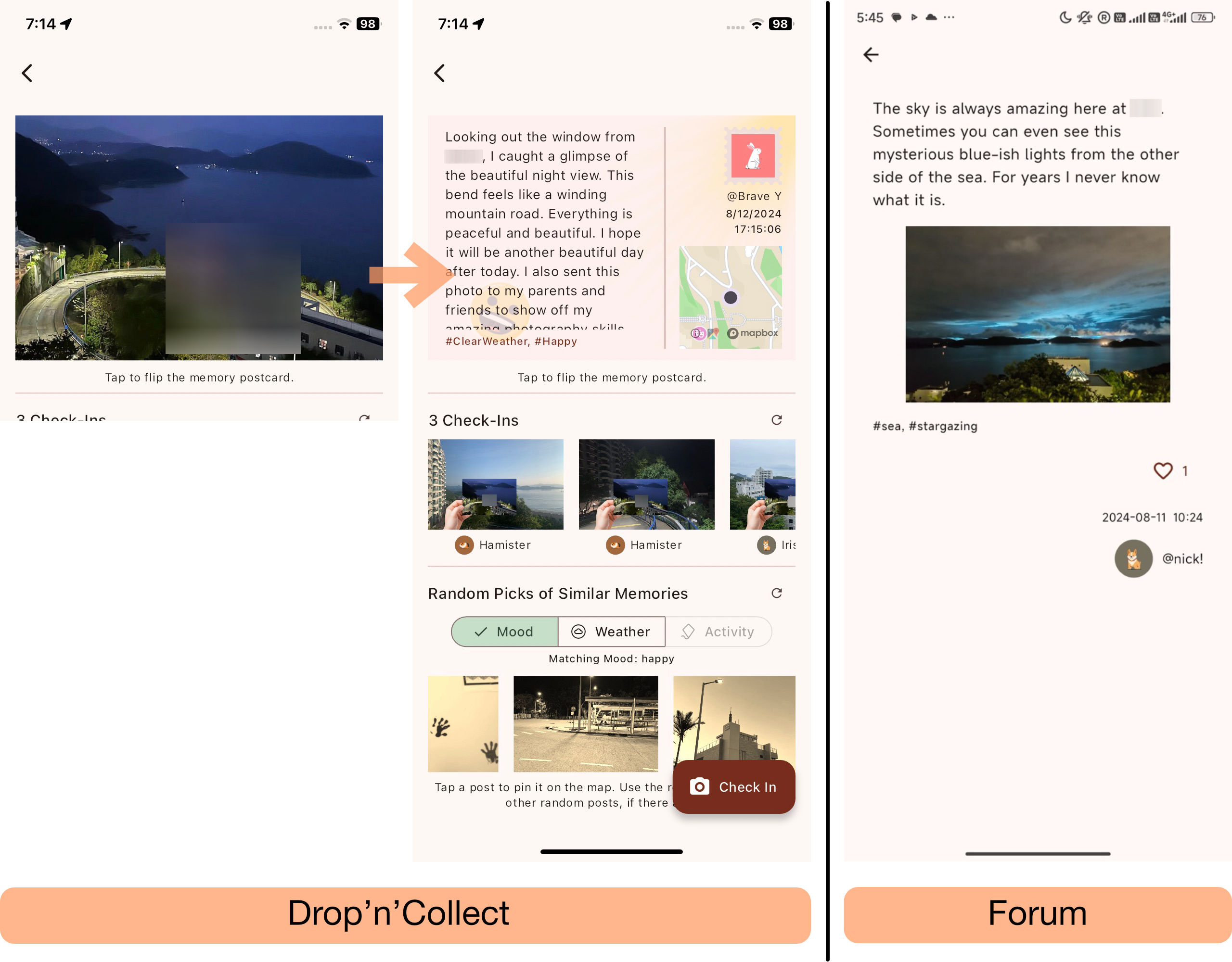}
    \caption{Viewing a memory post in both systems}\label{fig:view}
    \Description{Screenshots when viewing a memory post on two systems. For \name, a postcard on the top is flippable between the photo side and the content side. Below is a horizontal list of check-in photographs, each being a user's picture and an overlay of a fake hand holding the postcard. Further below is a horizontal list of sepia photos as the recommended memories. For the forum, vertically laid out are the text content, the photo, the hashtag list, the ``like'' button, the date and time of the post, and the author's ID.}
  \end{subcaptionblock}
  \hfill
  \begin{subcaptionblock}[T]{\textwidth}
    \centering\vspace*{0.2cm}
    \includegraphics[width=\textwidth]{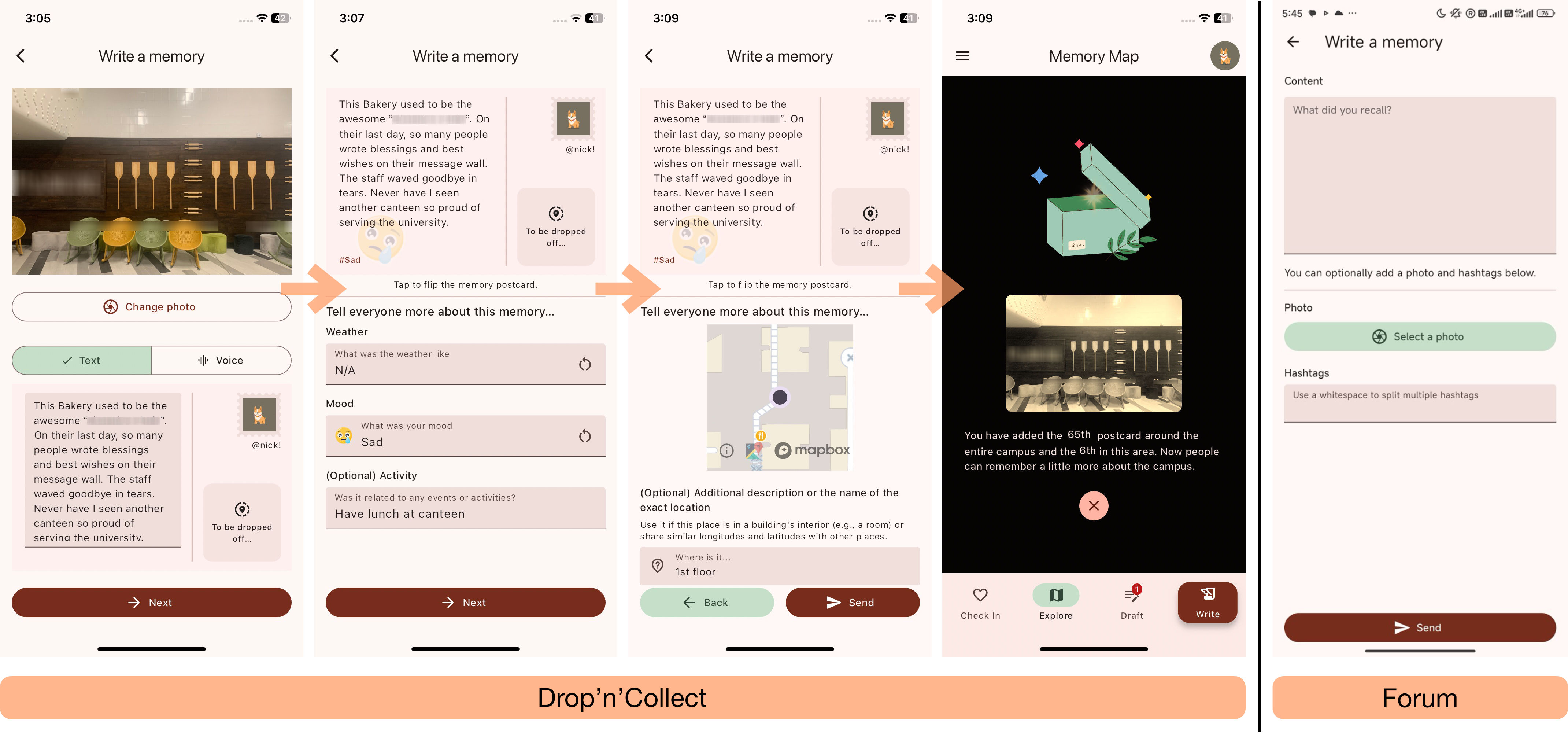}\vspace*{-0.2cm}
    \caption{Writing memory posts in both systems\vspace*{0.2cm}}\label{fig:write}
    \Description{On the left are four screenshots showing the procedure of publishing a memory postcard in \name{}: choose a photo, write or speak content on a virtual postcard, select emotions and weather, and optionally input an activity, confirm the location, and optionally clarify the location, see the thank-you message. On the right is one screenshot of publishing a memory post in the forum: write the text content, optionally add a photo, and optionally add hashtags.}
  \end{subcaptionblock}
  \begin{subcaptionblock}[T]{0.58\textwidth}
    \centering
    \includegraphics[width=0.32\textwidth]{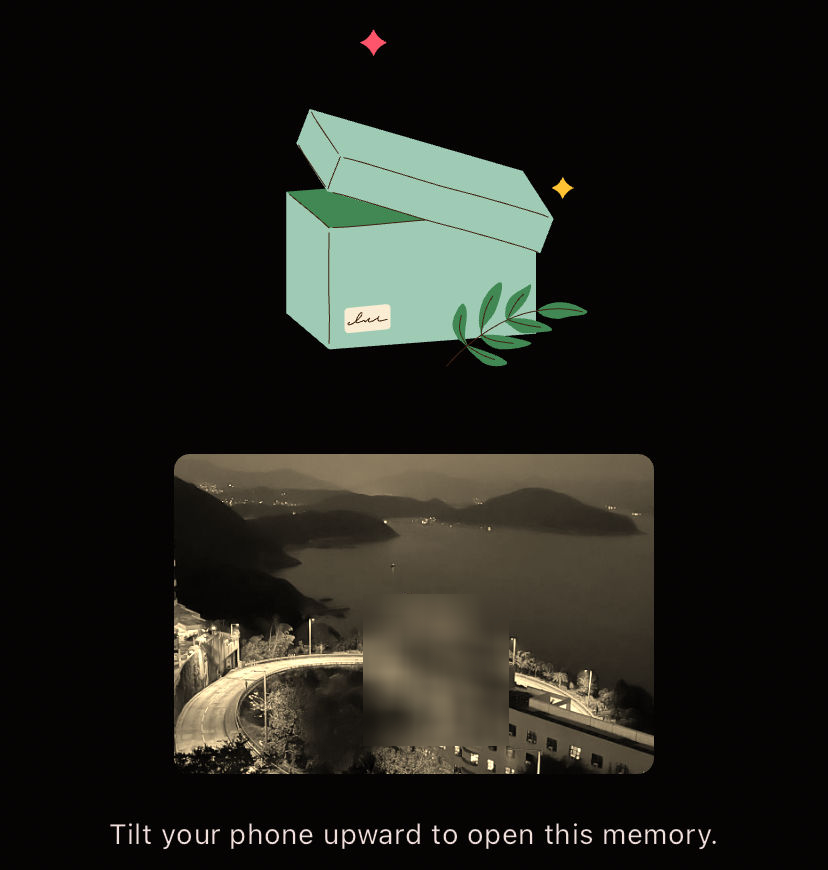}
    \includegraphics[width=0.32\textwidth]{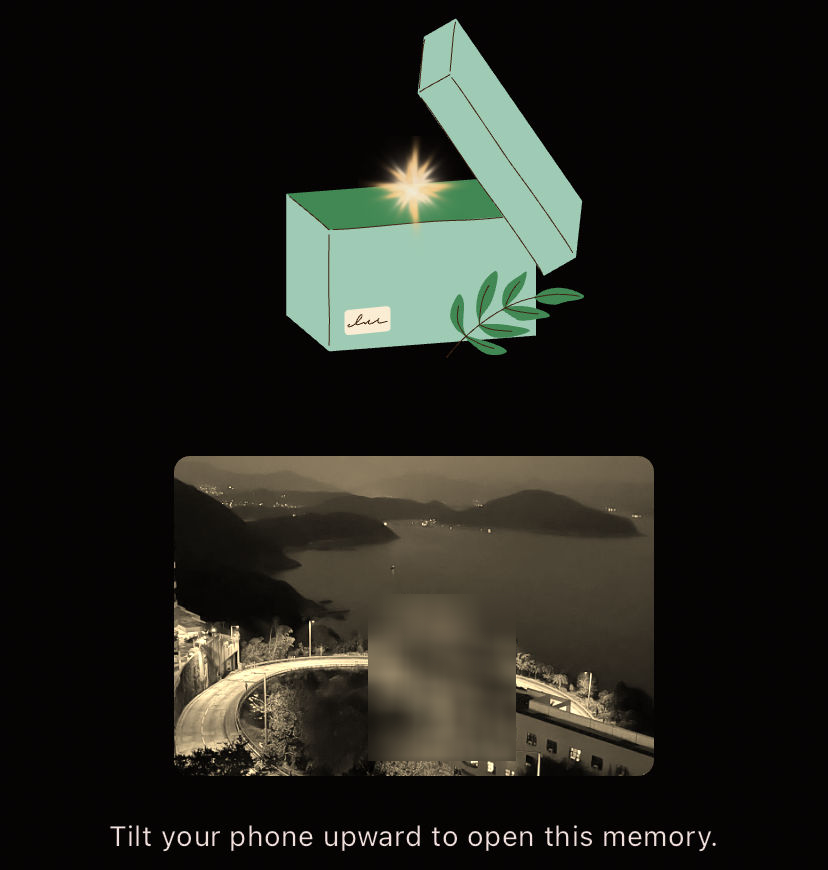}
    \includegraphics[width=0.32\textwidth]{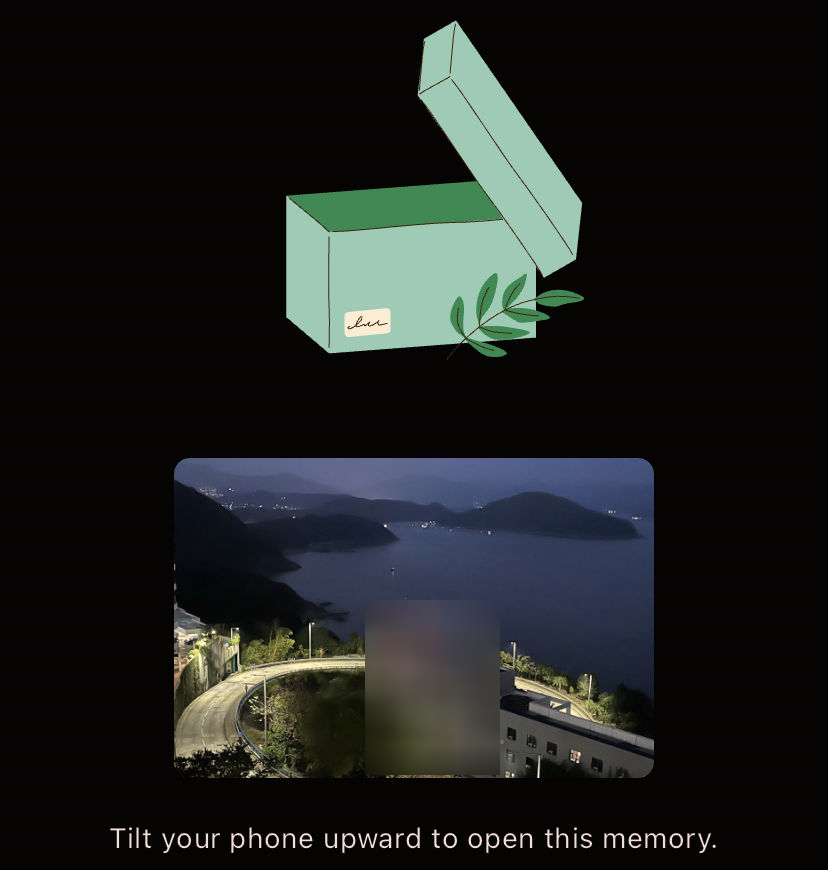}
    \caption{In \name, users are asked to tilt the mobile phone upward to open a memory post. It is to mimic the action of opening a memory box.}\label{fig:open}
    \Description{Three screenshots that show the following animation: a green box opens, a yellow-white sparkling item appears from inside the box and flies to a sepia photograph, and the sepia photograph then recovers its original color.}
  \end{subcaptionblock}
  \hfill
  \begin{subcaptionblock}[T]{0.4\textwidth}
    \centering
    \includegraphics[width=\textwidth]{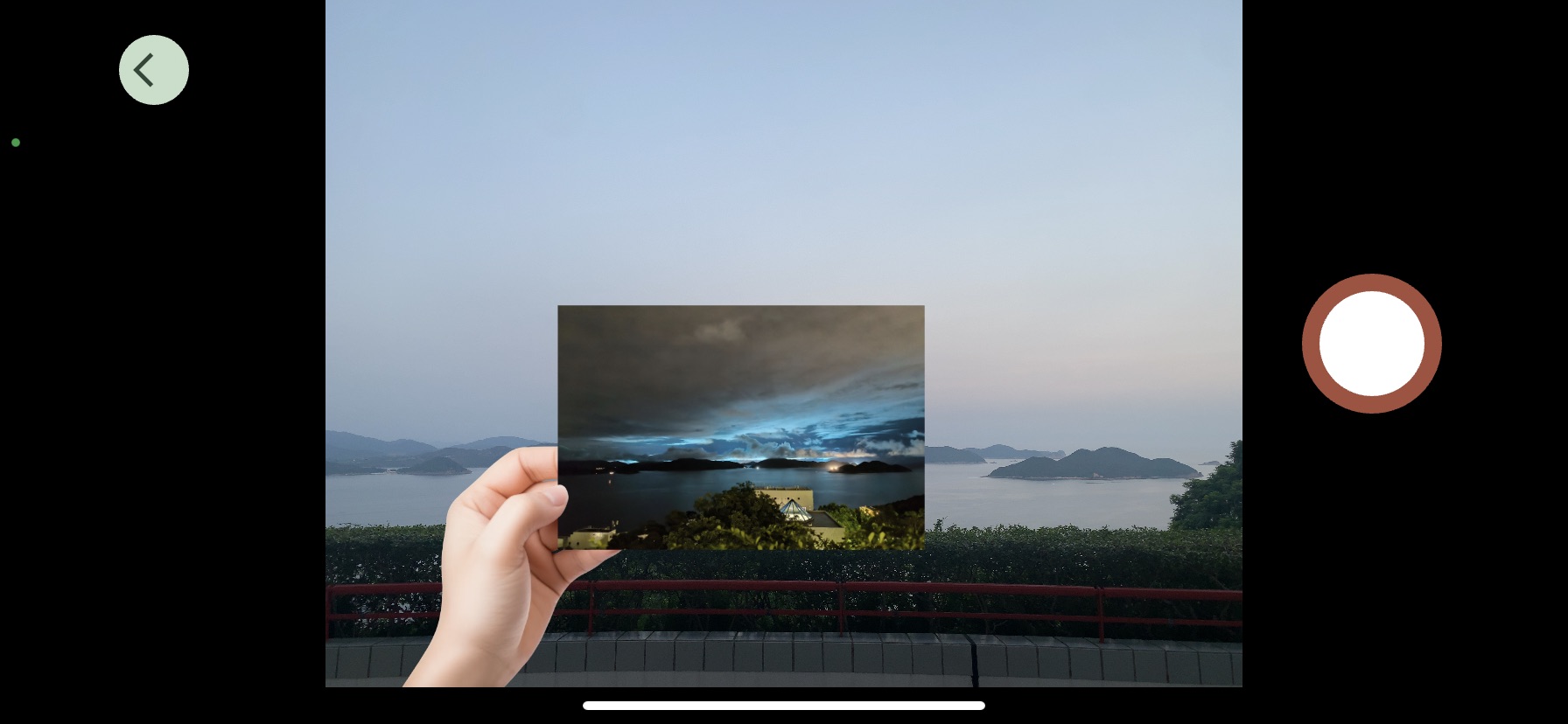}
    \caption{In \name, users check in a memory post with photography. A fake hand holding the memory postcard overlays the viewfinder and the photo.}\label{fig:check-in}
    \Description{A camera view with the overlay at the bottom center of a fake hand holding a postcard.}
  \end{subcaptionblock}
  \caption{More screenshots of the feature pairs in both systems (\ref{fig:browse}--\ref{fig:write}) and unique features in \name(\ref{fig:open}--\ref{fig:check-in}). The landmarks' names on the map are temporarily removed for anonymization.}\label{fig:screenshots}
\end{figure*}

In this section, we first explain the pairing features in both system and their rationales, grouped in several subsections. Then, in \autoref{sec:design:features:usages}, we describe both systems' complete usages and their technical implementation. The pairing feature sets are also organized into \autoref{tab:features}, and \autoref{fig:screenshots} shows several screenshots of both apps.

\begin{table*}[ht]
  \caption{Matching features in two systems. ``DnC'' stands for \name and ``FRM'' stands for forum.}\label{tab:features}
  \small
  \begin{tabular}{@{}p{2cm} >{\raggedright\arraybackslash}p{7.1cm} >{\raggedright\arraybackslash}p{4.1cm} @{}}
    \toprule
    \textbf{User Intent} & \textbf{Features} & \textbf{Hypothetical Effects} \\
    \midrule
    \multirow[t]{2}{\linewidth}{View memories}
    & DnC: Users can only see and open nearby memory posts. & DnC: Engaging, impressive \\
    & FRM: Users can browse and open all memory posts anywhere. & FRM: Accessible, comprehensive \\
    \midrule
    \multirow[t]{2}{\linewidth}{Post memories}
    & DnC: Users' memory posts are tied to the current geolocation. & DnC: Evocative, ritualistic \\
    & FRM: Users' memory posts are appended to the top of the feed. & FRM: Efficient \\
    \midrule
    \multirow[t]{2}{\linewidth}{Impression of time}
    & DnC: The UI conveys metaphors for postcards and memory boxes. & DnC: Impression of memory \\
    & FRM: The UI conveys no metaphors for time or history. & FRM: Simplicity \\
    \midrule
    \multirow[t]{2}{\linewidth}{Acknowledge others' memories}
    & DnC: Users check in the posts via a photograph. & DnC: Poetic, connecting readers of the same post \\
    & FRM: Users tap the like button. & FRM: Quick, informative \\
    \midrule
    \multirow[t]{2}{\linewidth}{Supplement memories with metadata}
    & DnC: Users can optionally add ``emotion'', ``weather'', and ``activity'' information, which is pre-filled by GenAI. & DnC: Specific \\
    & FRM: Users can optionally add free-form text hashtags. & FRM: Flexible \\
    \midrule
    \multirow[t]{2}{\linewidth}{Explore memories by metadata}
    & DnC: Users are recommended memories in other regions that share similar metadata. & DnC: Encouraging physical exploration \\
    & FRM: Users are not specifically recommended to explore similar posts far away. & FRM: In line with geolocation non-limit \\
    \bottomrule
  \end{tabular}
\end{table*}

\subsubsection{Memory Exploration and Creation}\label{sec:design:features:loc}

In \name, users' memory posts are tied to their current physical geolocations, and they can only see the memory posts nearby. This way, the exploration and curation of memories occurs in conjunction with physical exploration of the campus space, forming a locative narrative experience~\cite{nisiAuthoringLocativeNarratives2022}.
We intend to leverage the in-person witness of the physical surroundings to stimulate writers' recollection and augment readers' impression (\textbf{DR1}). And we also intend to encourage the process of physical exploration, which may add value to the memories users discover and read (\textbf{DR1}, \textbf{DR5})

Some refinements are made to smooth the locative narrative design.
Considering the potential inconvenience of writing text while walking around (\textbf{DR3}),
\name allows users to write first and ``drop off'' later, or save the current location first and write later. Users can also post audio memories and listen to text-to-speech (TTS) audios of text memory posts.
Then, as the GPS in our city is not very stable, it did not work well to directly center the visible area at the user's GPS location (as the content may not be really around them). Hence, we segmented geolocation regions around the entire campus and only checked which region the user was currently in. In this more tolerant way, all posts in the same region were visible to the user.
The regions were handcrafted through careful discussions. Two researchers first independently zoned the entire campus, following the criteria that each region contained at most one specialized building or area, served at most one purpose, and did not split any point of interest. Then, three researchers collaboratively discussed to settle the conflicts, additionally requiring every region's size between a tennis court and a track-and-field ground so that users could ``see through'' the region when they were physically there.

Meanwhile, the forum system adopts the most common infinite list of information feed. Users' new memory posts are directly appended to the top of the feed, and they can browse the entire community's memories without geolocation limits.
This way, we expect that the writing experience may be less evocative but more efficient; as for the reading experience, memories may be more accessible but less immersive.

\subsubsection{Visual Elements}

We propose two concrete skeuomorphic metaphors for \name: postcards and memory boxes.
Postcards are often used to hold text expressions and graphics about a particular experience~\cite{gerritsenMailingArchivedEmails2016} for meaningful reflection~\cite{sunPostcardYourFood2020} and sharing with others~\cite{kellyUnderstandingParticipationOpportunities2012}.
Each memory post is made into the shape of a postcard, with a required photograph as the cover, the memory's geolocation as the address, and the author's avatar as the stamp (\autoref{fig:view}).
Viewers can flip between the photo and content sides of the postcard.

Memory boxes afford to contain and transmit memory-related inventories~\cite{aaliMemoryBoxesExperimental2014} (\textbf{DR4}), and they afford the interactions of opening and closing~\cite{kleinbergerSupportingElderConnectedness2019},
which may be integrated into our system to increase interactivity and engagement (\textbf{DR5}).
The memory box is designed to be an interactive exercise of ``opening a post'': users tilt their mobile phones upward to mimic the action of opening a box.
We postulate that the tilting-up action may also guide the user to lift up the head, bringing the vision from the screen to the physical surroundings (\textbf{DR4}).
To link this metaphor to the postcard,
the memory box contains the ``spirit'' of a memory that refreshes the time-washed photograph back to normal.
Similarly, when users post a new memory, a reversed animation stores the ``spirit'' into the memory box.
A thank-you message also signifies the user's contribution to the community's collective memory (\textbf{DR3}).

For the forum system, neither of the metaphorical visual patterns mentioned above is applied. Its UI is intended to be neutral as a generic forum. Despite the less explicit association with memories, it may result in simpler and more efficient interactions.

\subsubsection{Interactions with Memory Posts}

A ``like'' button is commonly included in online platforms, as it allows lightweight acknowledgement of other users' outputs. In our forum system, we also retain this traditional feature. Concerning its matching feature in \name, we extend the metaphor of postcards mentioned above. Especially, we leveraged a real-world practice of photographing a postcard at the spot of its origin, which created an illusion of stitching the environment and the card together\footnote{An example of postcard photography is \url{https://www.instagram.com/filmtourismus/}. This account is a photography influencer and is not related to the authors' identities.}. Users could open the camera from any memory post's screen, where the postcard photo and a fake hand image were overlaid on the viewfinder and the final photo that users captured. Since \name users already arrive at the specific spot to consume a memory post, the photography indicates a ``checking-in'' with minimal extra physical exertion. It could offer poetic connections with a memory in the community, as well as other readers who have once shown up here (\textbf{DR4}).

Online forums usually offer the comment feature for more serious verbal interactions. We provide the comment feature as-is in both systems, considering the otherwise limited range of interactions with memory content. However, the comments are not the main focus of our current study. The feature only serves to complete UGC-driven platforms.

\subsubsection{Memory Metadata}\label{sec:design:features:metadata}

To complement the core elements of collective memory-related content (see \autoref{sec:design:approach}), we cater to the activities and personal contexts as posts' metadata in \name{}. Two specific aspects of personal contexts, emotions and weather, were chosen due to their usual association with the memory experience~\cite{sasDesigningCollectiveRemembering2006}.
The activity is free-form text, whereas the aspects of the personal context are in options. The emotions are from the basic emotion theory~\cite{anEmoWearExploringEmotional2024} (happy, angry, afraid, surprised, sad, calm, and neutral), and the weathers are from Apple's list of weather icons\footnote{\url{https://support.apple.com/guide/weather-mac/symbols-used-in-weather-apdw00197118/mac}}, excluding mixed weathers and those not available locally (clear, cloudy, rainy, windy, foggy, not applicable).
To further lower their cognitive load of additional input (\textbf{DR3}), we wrote simple GenAI prompts to pre-fill them based on the post content.

Apart from presenting these metadata tags, we recommend posts based on the similarities in either dimensions\footnote{The similarity of free-form text is calculated from their GenAI embeddings}, aiming to further raise users' awareness of collectivity (\textbf{DR2}).
Users can see five random recommendations beneath the postcard, and they can request another five random memories (\autoref{fig:view}).
All recommendations are outside the current region, encouraging users to step out and embrace serendipitous discoveries of diverse content and locations.

In the forum system, since we do not explicitly practice Lak et al.'s framework about collective memory's core elements~\cite{lakCollectiveMemoryUrban2019}, we do not offer metadata in specific dimensions. To provide a matching feature for the sake of informativeness, users can write free-form text as hashtags --- similar to many real-world forum designs --- beneath the post. It does not specifically recommend far-away posts, as users always browse all posts without geolocation limits. We expect that the metadata in the forum system may add less value compared to \name, because users may have already written more informative content, and they are already accessible to all the posts.

\edit{
  \subsubsection{System Usages}\label{sec:design:features:usages}

  In \name, users are greeted with the map view page shown in \autoref{fig:browse}.
  The map is always centered at the user's current geolocation, unless temporarily dragged aways manually by the user to see the nearby memory distributions.
  At any time, the map shows the current region the user is in and the exact locations of the memories in this region. At the bottom is a horizontally scrollable list of the photos attached to all the memories in this region.
  A photograph initially has the sepia filter, implying the sealed past; after the user has read this memory, it regains its original colors.

  As the user taps to open a memory postcard, the tilting interaction shown in \autoref{fig:open} pops up, and it proceeds to the viewing page (\autoref{fig:view}) after the user completes it.
  The same memory postcard is at the top of the viewing page, and the user can tap on it to flip between the photograph and text pages.
  If it is an audio memory, the text content area is replaced by the play, pause, and rewind buttons, and the playback progress.
  Below the postcard content is a horizontal list of all check-ins at this memory, the user can scroll to see how other people have left their traces of visits at different times or tap the top-right button to refresh the list. The user can also check in himself/herself using the floating button on this page, which opens up a camera page (\autoref{fig:check-in}).
  Further below is another horizontal list of memory recommendations. After the user selects a metadata type, at most five random memories in other regions with similar metadata are recommended (detailed in \autoref{sec:design:features:metadata}). The user can also refresh the list to get some other random memories.
  As the user taps one of the recommendations, a marker tells the user that the location of this memory is pinned on the map. The user can return to the map view and head for it, or retap to cancel. Upon arriving at the target memory's region, the same memory postcard is also highlighted in the regional list below the map.

  Using the bottom tabs at the bottom of the map view, the user can revisit the list of all memory postcards once checked in (without going to their geolocations again), write a new memory, or revisit the list of memory drafts and complete any of them.
  The writing process starts with a choice to write up a memory right here, write first and drop off later, or save the geolocation and write later. Then, with adjusted orders, it involves uploading a photograph, completing the textfield or an recording prompt, filling in the metadata, and confirming the geolocation (\autoref{fig:write}).
  Upon completing a memory (either writing up a memory at once or finishing a draft), it displays the thank-you animation to acknowledgement the contribution, and returns to the map view.

  In the forum system, the user starts with a list view of all memory posts' previews regardless of geolocations, sorted by their post dates and infinitely scrollable (\autoref{fig:browse}).
  The user taps any entry to browse its full content on a dedicated page without (\autoref{fig:view}). The heart toggle button allows the user to like this memory and see how many users have liked it.
  Using the bottom tabs of the list view, the user can also go to the list of all memories once liked, as well as write a new memory post.
  The writing page consists of the content, a photograph, and user-defined, free-text hashtags (\autoref{fig:write}).
  There are no additional animations upon entering or publishing a memory post.

  \subsubsection{Technical Implementation}\label{sec:design:features:implementation}
  Regarding the technical implementation, both systems' frontends\footnote{Open-sourced at \anon{\url{https://github.com/fhfuih/Campus-Collective-Memory-Frontend}}.} were implemented as native cross-platform mobile apps in Flutter, with the same UI style using Figma and the Material Design framework. Data storage and manipulation of either app were supported by a dedicated backend\footnote{Open-sourced at \anon{\url{https://github.com/fhfuih/Campus-Collective-Memory-Backend}}.} in Go and PostgreSQL database, all hosted on a private server.
  We distribute the iOS versions of both systems as TestFlight\footnote{\url{https://testflight.apple.com/}, the official and only way to distribute iOS apps without paying for a seat on App Store.} beta apps and the Android versions as \texttt{.apk} installer files, both delivered with installation tutorials and system instructions.
}

\section{Evaluation}

To answer the research questions, we conducted an \edit{exploratory}, between-subjects field study.
\edit{With the two design probes described above,} we aimed to uncover community members' grassroots interpretations of collective memory and its co-contribution process.

\edit{
  \subsection{Scenario}\label{sec:eval:scenario}
  Among existing literature, we have identified two primary scenarios when investigating a community's collective memory.
  Some research focuses on the scenario of users' explicit and purposeful acts that contribute to collective memory, as well as the possible use of various intentional designs for collective memory development~\cite{jonesCoconstructingFamilyMemory2018,lindleyNarrativeMemoryPractice2009,sasDesigningCollectiveRemembering2006,jonesDesigningSocialMemory2016}.
  Meanwhile, many other studies adopt the case where collective memory is an organic by-product on general-purpose platforms (e.g., Wikipedia~\cite{ferronCollectiveMemoryBuilding2011,twymanBlackLivesMatter2017}, social media~\cite{aldamenStimulationCollectiveMemory2024,zhangCrisisCollectiveMemory2020,khlevnyukNarrowcastingCollectiveMemory2019}), discussing its possible integration into people's generic online social activities.

  In our study, the bottom-up approach required us to directly probe participants' own conceptualizations of the term.
  Thus, we applied the first scenario, explicitly asking participants to contribute to collective memory.
  By making memory-making the primary goal, we could better investigate how non-experts interpret and act upon this abstract concept, ultimately providing human-centered design insights.
}

\subsection{Methods}\label{sec:eval:methods}

Semi-structured interviews were our primary means to investigate participants' experiences and reflections. Our questions covered participants' understanding of their ongoing procedure of building collective memory, what content was counted as its constituents, whether their experiences triggered further reflection on the conceptualization of collective memory, and how the process could be supported by specific features or affordances. After recalling from their own experiences, participants are also shown the content on the other platform for any comparative comments.

\begin{figure}[p]
  \begin{sideways}
    \begin{minipage}[c][\textwidth][c]{\textheight}
      \includegraphics[%
        width=\linewidth,%
        trim=5cm 5cm 5cm 5cm,%
      ]{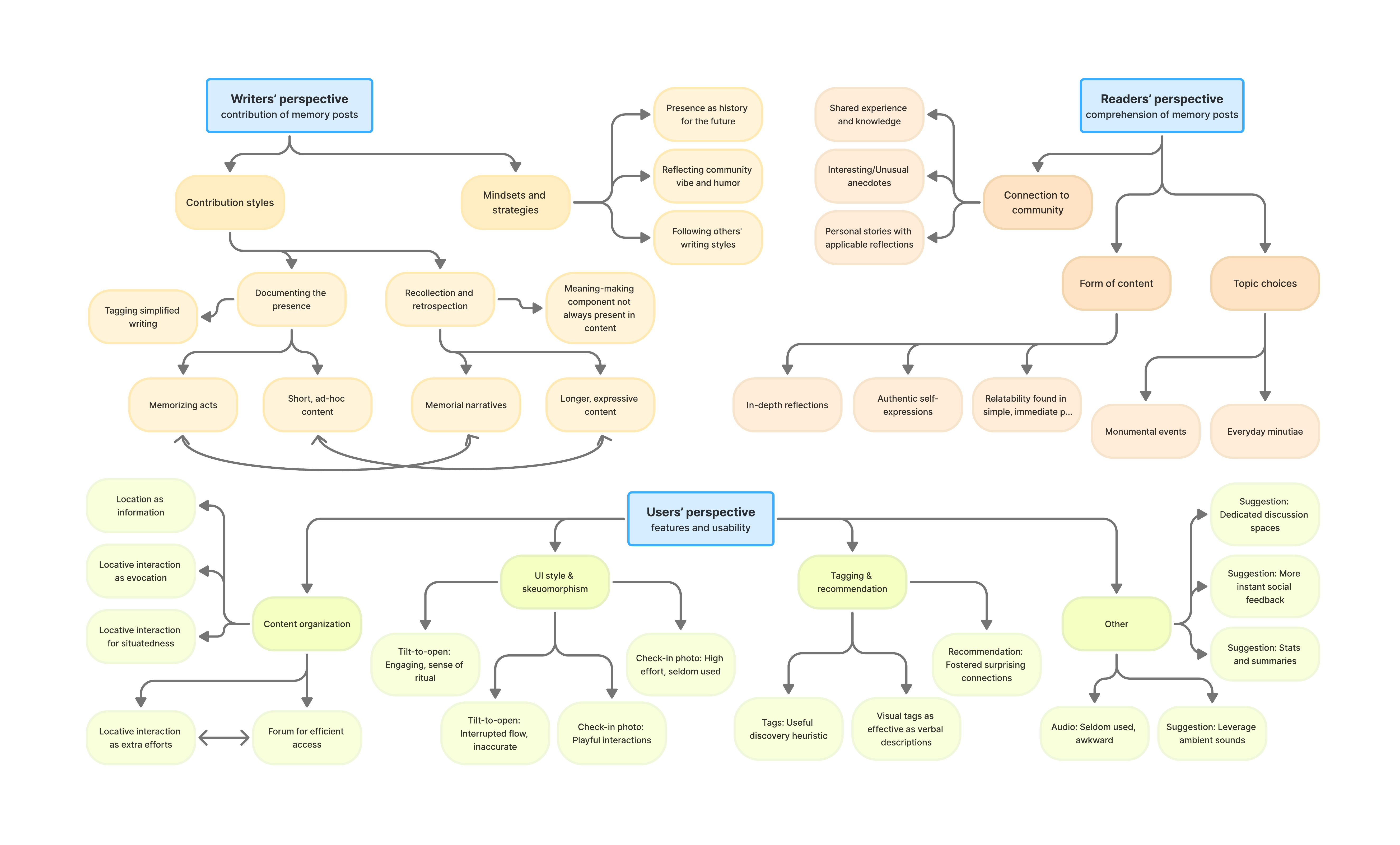}
      \caption{\edit{The thematic map of our qualitative findings.}}\label{fig:codebook}
    \end{minipage}
  \end{sideways}
\end{figure}

\edit{The interview transcripts underwent a thematic analysis.
  We started deductively with three high-level themes of user interactions relevant to our research questions and interview structure~\cite{braunThematicAnalysisPractical2022}: participants' contribution of memory content (\ref{rq:conceptualization}, \ref{rq:practice}), their reading and comprehension of others' content (\ref{rq:conceptualization}, \ref{rq:practice}), and feature usability (\ref{rq:practice}, \ref{rq:considerations}).
  Guided by the three themes, we then inductively extracted codes to capture the emerging insights~\cite{braunThematicAnalysisPractical2022}, such as the characteristics of user posts, their interactional patterns, and users' mental interpretations.
}
During the coding, two researchers independently coded the data first. Two subsequent rounds of joint discussions resolved the conflicts and confusions between similar codes across themes.
\edit{
  The final codes are presented in a thematic map~\cite{ahmedUsingThematicAnalysis2025} (\autoref{fig:codebook}), and elaborated in the following section.
}

The questionnaires were to supplement our understanding of the user feedback. In line with the major benefit of collective memory, we inspected whether the systems could facilitate connectedness with community members~\cite{sasDesigningCollectiveRemembering2006} with the Campus Connected Scale (CCS)~\cite{leeCulturalOrientationMulticultural2000} and the collective identity~\cite{roedigerCollectiveMemoryNew2015,eyermanPresentCultureTransmission2004} with Collective Self-Esteem Scale (CSES)~\cite{luhtanenCollectiveSelfEsteemScale1992}.
(We excluded the ``membership'' and ``public'' sections of CSES. The first overlapped with CCS. The second is about the public image of being part of the group, and was out of this study's scope.)
They were sampled before and after the experience period, and we calculated the difference between the two samples.
In addition, we used UEQ-S~\cite{schreppDesignEvaluationShort2017} plus feature-specific customized questions only in the exit questionnaire to see how the current design supported the practical co-contribution of collective memory.

\edit{
  We chose the Mann-Whitney U test~\cite{mannTestWhetherOne1947} to compare the questionnaire data (i.e., the increments in CCS and CSES items, the direct responses to other items) for the following reasons.
  We first used Shapiro-Wilk normality tests~\cite{ghasemiNormalityTestsStatistical2012} to verify that either group's responses to most questions did not follow normal distributions. (Only two out of 44 questions had both group's responses in normal distributions, so we still considered the overall data not normal.)
  This combined with the small sample size (below 50) suggested nonparametric tests~\cite{krzywinskiNonparametricTests2014,kapteinPowerfulConsistentAnalysis2010}.
  Furthermore, different individuals divided in two groups form independent samples, for which the Mann-Whitney U test is the most common nonparametric test method~\cite{nacharMannWhitneyTestAssessing2008}.
  Prior HCI scholars also validate and recommend the Mann-Whitney U test for Likert scale data in common HCI studies~\cite{kapteinPowerfulConsistentAnalysis2010}.
}

Lastly, we also did a post-hoc content analysis on users' memory posts to see the contents' traits in both apps. We first counted the number of words in each post as an overview. Then, we identified high-level aspects that emerged in their posts as well as lower-level topics that their content was about. Two researchers independently coded the content after reading through every post, and another researcher joined the discussion to resolve the conflicts. By counting their distributions, we wanted to better capture participants' understanding of the constituents of collective memory, and whether different feature designs affected the content they would contribute.

Again, although much comparison was involved, we did not aim to see whether one system design is better than the other.
Rather, we sought to uncover users' perceptions and gather heuristics for the design space through juxtaposing these two design instances.

\subsection{Participants}

We recruited 40 participants (24 male, 15 female, 1 undisclosed) from the local university we deployed the system. Participants were recruited through word of mouth and social media. They were equally and randomly divided into two groups of 20, and labeled A1--A20 (for \name) and B1--B20 (for forum).
One participant in each group quit midway, resulting in 19 participants each in the final evaluation. Their demographic information is listed in \autoref{tab:demog}.

\edit{
  While we were sampling and grouping participants, we ensured that each group had both newcomers to the university and experienced residents, so that they were more likely to have both memory sharing and discovery.
  Thus, we added a self-reported ``year (of residence) on campus'' metric in the recruitment survey.
  As we grouped the participants, we ensured that this attribute followed similar distributions in both groups (shown in \autoref{tab:demog}). Group A has a mean of 3.16 and std of 2.58, and Group B has a mean of 2.79 and std of 2.44. A further Mann-Whitney U test on this attribute also verified statistical insignificance (\(p=0.801\)). (Similar to \autoref{sec:eval:methods}, we used the Shapiro-Wilk test~\cite{ghasemiNormalityTestsStatistical2012} to verify non-normality and followed the same subsequent rationale to choose the Mann-Whitney U test.)
}

\begin{table}[ht]
  \caption{Participants' demographic backgrounds. ``Years on campus'' are rounded down.}\label{tab:demog}
  \scriptsize
  \begin{tabular}{@{}lllp{1cm}ll@{}}
    \toprule
    \textbf{ID} & \textbf{Age} & \textbf{Gender} & \textbf{Years on campus} & \textbf{Identity*} & \textbf{Interview} \\
    \midrule
    A1  & 24 & Female      & 6 & RPG    & Y \\
    A2  & 25 & Female      & 7 & RPG    & Y \\
    A3  & 26 & Male        & 1 & RPG    & Y \\
    A4  & 24 & Male        & 2 & RPG    & Y \\
    A5  & 25 & Male        & 8 & Alumni & Y \\
    A6  & 24 & Female      & 3 & RPG    & Y \\
    A7  & 23 & Female      & 1 & TPG    & Y \\
    A8  & 22 & Undisclosed & 5 & RPG    & Y \\
    A9  & 22 & Male        & 0 & RPG    & Y \\
    A10 & 23 & Male        & 4 & RPG    & Y \\
    A11 & 24 & Female      & 0 & RPG    &   \\
    A12 & 21 & Male        & 3 & UG     &   \\
    A13 & 27 & Male        & 5 & RPG    &   \\
    A14 & 24 & Male        & 6 & RPG    &   \\
    A15 & 23 & Male        & 0 & RPG    &   \\
    A16 & 21 & Female      & 0 & RPG    &   \\
    A17 & 26 & Female      & 4 & RPG    &   \\
    A18 & 23 & Male        & 0 & TPG    &   \\
    A19 & 26 & Female      & 5 & RPG    &   \\
    \bottomrule
  \end{tabular}
  \hfill
  \begin{tabular}{@{}lllp{1cm}ll@{}}
    \toprule
    \textbf{ID} & \textbf{Age} & \textbf{Gender} & \textbf{Years on campus} & \textbf{Identity*} & \textbf{Interview} \\
    \midrule
    B1  & 24 & Male   & 2 & RPG   & Y \\
    B2  & 27 & Male   & 3 & RPG   & Y \\
    B3  & 24 & Female & 6 & RPG   & Y \\
    B4  & 25 & Male   & 3 & RPG   & Y \\
    B5  & 29 & Female & 4 & RPG   & Y \\
    B6  & 25 & Male   & 6 & RPG   & Y \\
    B7  & 18 & Female & 0 & UG    & Y \\
    B8  & 33 & Male   & 1 & Staff &   \\
    B9  & 20 & Female & 1 & UG    &   \\
    B10 & 30 & Male   & 1 & Staff &   \\
    B11 & 19 & Female & 1 & UG    &   \\
    B12 & 25 & Male   & 7 & RPG   &   \\
    B13 & 23 & Male   & 5 & RPG   &   \\
    B14 & 29 & Female & 1 & Staff &   \\
    B15 & 24 & Male   & 6 & RPG   &   \\
    B16 & 25 & Male   & 0 & TPG   &   \\
    B17 & 24 & Male   & 6 & RPG   &   \\
    B18 & 21 & Male   & 0 & V     &   \\
    B19 & 26 & Male   & 0 & RPG   &   \\
    \bottomrule
  \end{tabular}

  \vspace{0.5em}
  \begin{minipage}{\linewidth}
    * UG = Undergraduate;\enspace RPG = Research Postgraduate;\enspace TPG = Taught Postgraduate;\enspace V = Visiting Student
  \end{minipage}
\end{table}

\subsection{Procedure}\label{res:procedure}

Before the evaluation, we inserted the same 23 memories as seed content to bootstrap engagement at the beginning of the evaluation~\cite{cranshawJourneysNotesDesigning2016}.
They were authentic memories of three researchers and three external authors recruited via on-campus advertisement.
We deliberately recorded memories of different places on campus and balanced text and audio posts. The same seed content was stored in each system's database, and the audio posts were transcribed into text in the forum system.

We allocated a fixed two-week period for both groups to freely explore the system and contribute content that they believed was meaningful for the campus community's collective memory.
\edit{The experience period intentionally incorporated a final exam week, which is a typical recurring event of campus life.}
Before the two-week period starts, participants were first introduced to the study's purpose and the system's features, and they completed a preliminary questionnaire.
And during the period, they were allowed to engage with the system at any time and frequency that fitted their everyday lives.

After they finished the two-week session, they were invited to a 30-minute interview and another questionnaire.
Each participant received a 150 HKD stipend for completing the entire study procedure, with a 50 HKD deduction if dropping out of the interview.
There were 10 and 7 participants completing the interview (A1--A10, B1--B7, \autoref{tab:demog}).

\section{Results}

This section covers the study findings.
The findings are reported in three parts: participants' memory contribution, memory comprehension, and system usability.

\begin{table}[htp]
  \caption{The Mann-Whitney U tests of questionnaire ratings. The power is only reported if the result is significant (\(p < 0.05\)).}\label{tab:survey}
  \scriptsize
  \begin{tabular}{ >{\raggedright\arraybackslash}p{6.87cm} l l p{0.4cm} l >{\raggedright\arraybackslash}p{1.3cm} >{\raggedright\arraybackslash}p{1.1cm} }
    \toprule
    Question & U Stat. & Sig. & Effect Size & Power & {Mean (SD)\newline\name} & {Mean (SD)\newline Forum} \\
    \midrule
    There are people on campus with whom I feel a close bond. & 174.0 & \edit{0.851} & 0.488 & --- & -0.11 (1.24) & -0.16 (1.21) \\
    I feel that I belong around the people that I know on campus.* & 160.5 & 0.556 & 0.470 & --- & 0.26 (1.56) & 0.11 (1.45) \\
    I feel that I can share personal concerns with other people on campus. & 164.5 & 0.636 & 0.463 & --- & 0.53 (1.22) & 0.37 (1.21) \\
    I am able to make connections with a diverse group of people. & 202.0 & 0.499 & 0.547 & --- & 0.16 (0.83) & 0.32 (1.06) \\
    I don't feel distant from the other people on campus.* & 144.5 & 0.272 & 0.376 & --- & 0.68 (1.00) & 0.26 (0.87) \\
    I have a sense of togetherness with my peers.* & 179.0 & 0.975 & 0.500 & --- & 0.26 (0.87) & 0.26 (0.81) \\
    I can relate to my fellow schoolmates. & 140.5 & 0.217 & 0.358 & --- & 0.58 (1.07) & 0.11 (0.74) \\
    I don't lose my sense of connectedness with college life.* & 173.0 & 0.825 & 0.484 & --- & 0.21 (0.92) & 0.16 (0.90) \\
    I feel that I fit right in on campus. & 196.5 & 0.599 & 0.557 & --- & 0.05 (0.85) & 0.21 (0.71) \\
    I have a sense of brother/sisterhood with my friends at the university.* & 187.5 & 0.842 & 0.526 & --- & 0.63 (1.12) & 0.74 (1.19) \\
    I feel related to someone on campus.* & 199.5 & 0.562 & 0.576 & --- & 0.21 (0.79) & 0.47 (1.12) \\
    Other people make me feel at home on campus. & 209.0 & 0.378 & 0.549 & --- & 0.37 (1.12) & 0.53 (0.61) \\
    I don't feel disconnected from campus life.* & 176.5 & 0.913 & 0.468 & --- & 0.42 (0.96) & 0.32 (0.89) \\
    There is someone or some group that I feel I participate with.* & 162.5 & 0.561 & 0.417 & --- & 0.53 (1.07) & 0.26 (0.65) \\
    \midrule
    I don't regret that I belong to the campus community.* & 180.5 & 1.000 & 0.573 & --- & -0.11 (1.24) & 0.21 (1.18) \\
    In general, I'm glad to be a member of the campus community. & 204.0 & 0.459 & 0.562 & --- & 0.16 (0.83) & 0.32 (0.58) \\
    I feel that the campus community, which I am a member of, is worthwhile.* & 162.0 & 0.580 & 0.438 & --- & 0.21 (1.36) & -0.11 (1.49) \\
    I feel good about the campus community. & 168.5 & 0.721 & 0.469 & --- & 0.58 (1.07) & 0.47 (0.84) \\
    My membership in the campus community has much to do with how I feel about myself.* & 176.0 & 0.900 & 0.547 & --- & 0.26 (0.99) & 0.47 (1.47) \\
    The campus community is an important reflection of who I am. & 244.0 & 0.051 & 0.593 & --- & 0.21 (1.55) & 0.63 (0.90) \\
    The campus community is important to my sense of what kind of person I am.* & 163.5 & 0.619 & 0.432 & --- & 0.74 (1.28) & 0.32 (2.08) \\
    Belonging to the campus community is an important part of my self-image. & 178.0 & 0.952 & 0.500 & --- & 0.53 (0.96) & 0.53 (1.39) \\
    \midrule
    When using this platform, I can think of many content to share. & 130.5 & 0.128 & 0.369 & --- & 3.84 (1.34) & 3.21 (1.32) \\
    I think that the content I share is worth remembering by others. & 191.0 & 0.763 & 0.543 & --- & 3.42 (1.35) & 3.63 (1.38) \\
    I think that the content I share reflects something about the campus community (e.g., our history, our specialties, our daily lifestyles\dots). & 170.5 & 0.763 & 0.482 & --- & 4.58 (0.77) & 4.53 (0.84) \\
    I believe that my thoughts can be fully delivered, understood, and empathized with. & 130.0 & 0.134 & 0.352 & --- & 4.00 (1.25) & 3.32 (1.29) \\
    I believe that my sharing contributes to the community's collective memory. & 154.0 & 0.426 & 0.421 & --- & 4.37 (1.07) & 4.05 (1.18) \\
    I often clearly remember what other users have shared. & 204.0 & 0.488 & 0.545 & --- & 3.37 (1.26) & 3.58 (1.39) \\
    Other users' sharing that I have read are related to the campus community. & 208.5 & 0.388 & 0.604 & --- & 4.47 (0.96) & 4.79 (0.71) \\
    I can often understand and empathize with other users' sharing. & 205.0 & 0.457 & 0.560 & --- & 3.95 (0.97) & 4.16 (1.01) \\
    I am often impressed by other users' sharing. & 194.0 & 0.696 & 0.532 & --- & 3.58 (1.35) & 3.74 (1.41) \\
    Reading other users' sharing reminds me of my own memories to share. & 188.5 & 0.817 & 0.514 & --- & 4.05 (1.08) & 4.11 (0.99) \\
    Reading other users' sharing motivates me to share more of myself. & 133.5 & 0.156 & 0.344 & --- & 4.42 (1.07) & 3.68 (1.49) \\
    The content on this platform is organized in a logical way. & 70.0 & 0.001 & 0.207 & 0.095 & 4.84 (1.01) & 3.32 (1.57) \\
    I can navigate through the content on this platform in a logical way. & 75.0 & 0.001 & 0.213 & 0.098 & 4.79 (1.03) & 3.37 (1.46) \\
    While reading and writing on this platform, I pay balanced attention to not only the screen but the real-world campus environment as well. & 110.5 & 0.035 & 0.282 & 0.135 & 4.37 (0.90) & 3.26 (1.69) \\
    \midrule
    The system is obstructive/supportive & 155.0 & 0.384 & 0.425 & --- & 3.00 (0.47) & 2.84 (0.69) \\
    The system is complicated/easy & 224.0 & 0.158 & 0.626 & --- & 3.05 (0.71) & 3.37 (0.68) \\
    The system is inefficient/efficient & 154.0 & 0.404 & 0.423 & --- & 2.84 (0.83) & 2.63 (0.68) \\
    The system is confusing/clear & 190.5 & 0.767 & 0.547 & --- & 2.89 (1.05) & 3.05 (0.85) \\
    The system is boring/exciting & 119.5 & 0.055 & 0.322 & --- & 2.63 (0.60) & 2.16 (0.83) \\
    The system is uninteresting/interesting & 118.0 & 0.049 & 0.317 & 0.158 & 3.00 (0.75) & 2.42 (0.96) \\
    The system is conventional/inventive & 75.5 & 0.001 & 0.185 & 0.086 & 2.84 (0.76) & 1.63 (1.12) \\
    The system is usual/leading edge & \edit{85.0} & 0.003 & 0.225 & 0.103 & 2.53 (0.77) & 1.63 (0.90) \\
    \bottomrule
  \end{tabular}
  \begin{minipage}{\linewidth}
    *: These questions were negated in our survey to stay consistent with the original literature~\cite{leeCulturalOrientationMulticultural2000,luhtanenCollectiveSelfEsteemScale1992}. For the sake of presentational clarity, we reverse the question text and mirror the ratings in this table, so that larger numbers are always better.
  \end{minipage}
\end{table}

\subsection{Participants' Contribution of Memory Posts}\label{res:trait}

We first looked into how users posted memory posts in both systems, which most directly reflected their conception of co-contributing collective memory.
At first sight, \name users created significantly more posts that denoted the ``presence''---anything that users immediately saw, did, or thought of.
Consequently, there were more short contents like ``\textit{Late night working} [with a photo of a desk]'', ``\textit{Rainy\textasciitilde} [with a photo of an open-air plaza]'', and ``\textit{go swimming} [with a photo of the way to the swimming pool]''.
Interestingly, A8 saw a basketball with headphones attached to it on a nearby desk, and posted a picture of it with the caption ``What a handsome basketball!''
On the contrary, the forum posts were generally longer, more topically diverse, and there were more personal recollections related to the locations. For example, B7 wrote ``\textit{Staring at the horizon from the library. Is the breeze more bitter than the schoolwork? The scenery was the reason I chose this university without hesitation. Will I love the scenery as now after several years?}''
\edit{According to our observations, these memory outputs in the forum system present similar styles to those reported by other studies of collective memory on general-purpose social platforms: detailed, contextual, sometimes emotionally charged or polemical~\cite{twymanBlackLivesMatter2017,glawionRememberingWorldWar2023,zhangCrisisCollectiveMemory2020}.}

In the interview, participants mentioned that the locative design encouraged them to record interesting items or scenes immediately after they saw them (A3, A7), as well as their ad-hoc comments on the community's shared environments (A5).
As A3 mentioned, what he found meaningful about \name was that \textit{``it is more attached to people's everyday life\dots It encourages me to jot down something about this very moment at this location.''}
The tagging feature also conveniently replaced some participants' mental efforts of organizing languages (A1, A5, A7, A9), which potentially led to shorter memory text.
\edit{It is worth noting that, despite these different posting styles, there was no significant difference between the groups in their self-reported ability to conceive content to share (\(U=130.5\), \(p=0.128\), effect size 0.369). This suggests both systems provided adequate inspiration, albeit for different forms of expression.}
Concerning the longer, more personal and expressive content style in the forum system, this was partly because of the flexible timing of writing.
But more than that, the easy access to all posts sometimes also implicitly nudged participants to follow the writing style of previous posts.
As a result, the memories about the past were often meticulously selected and written expressively (B1, B7).
A similar phenomenon of following each other's writing styles was also spotted within \name users (A2, A3, A8, A9), and A7 especially expressed that \textit{``I could have written longer memories with more information like the other condition, but maybe because the posts I have seen are mostly short---I don't want to stand out too much.''}
\edit{Interestingly, even though the forum posts were longer on average, forum users did not feel significantly more capable of having their thoughts fully delivered and empathized with compared to \name users (\(U=130.0\), \(p=0.134\), effect size 0.352).} This finding challenges our initial expectation that longer, more detailed posts would be perceived as more effective for communication.

\begin{figure}[ht]
  \centering
  \begin{subcaptionblock}[t]{0.32\textwidth}
    \centering
    \includegraphics[width=\linewidth]{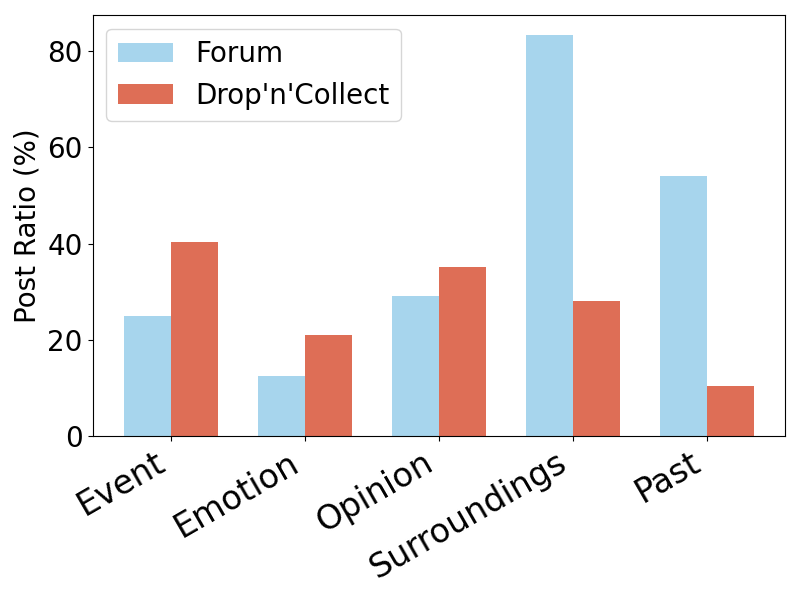}
    \caption{The proportions of posts that mentioned each aspect}\label{fig:content:percent-aspect}
  \end{subcaptionblock}
  \hfill
  \begin{subcaptionblock}[t]{0.32\textwidth}
    \centering
    \includegraphics[width=\linewidth]{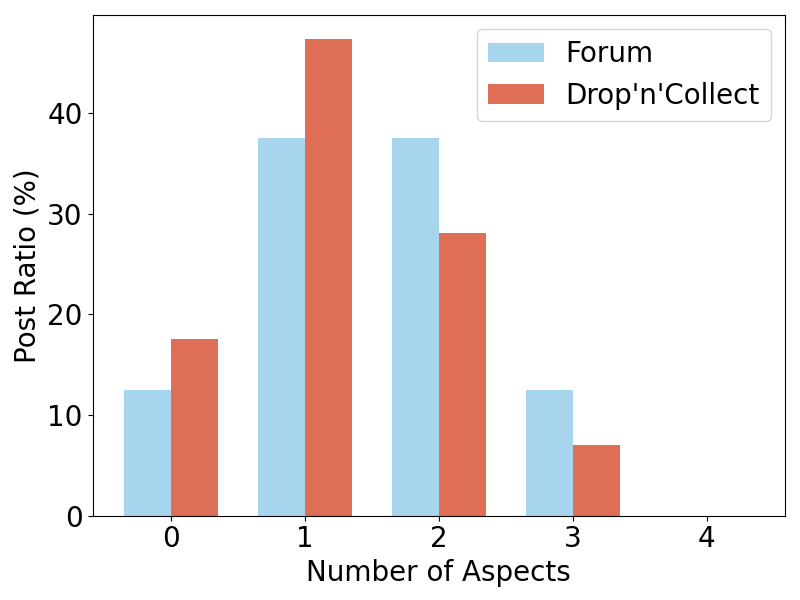}
    \caption{The histogram of the number of aspects mentioned in each post}\label{fig:content:n-aspect}
  \end{subcaptionblock}
  \hfill
  \begin{subcaptionblock}[t]{0.32\textwidth}
    \centering
    \includegraphics[width=\linewidth]{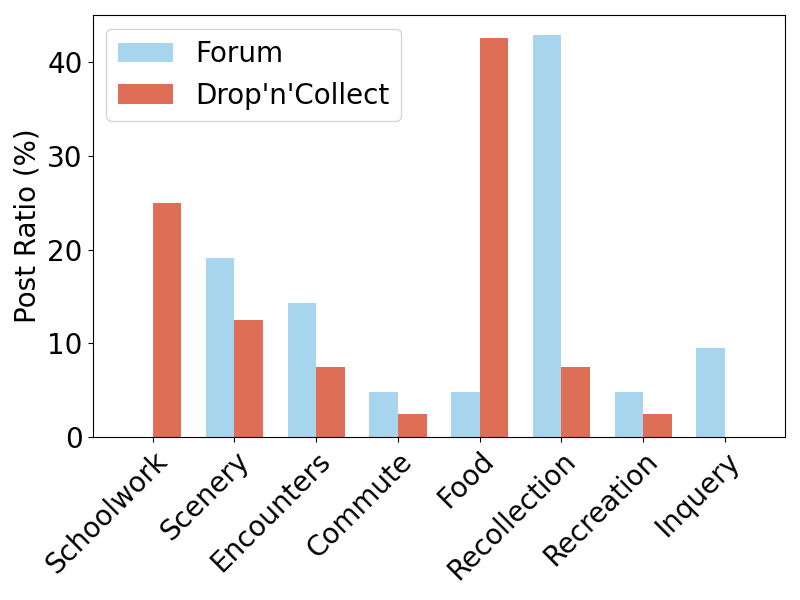}
    \caption{The proportions of posts that belonged to each topic}\label{fig:content:topic}
  \end{subcaptionblock}
  \caption{Aspects covered in each condition's posts.}\label{fig:content}
  \Description{(a) The forum has considerably more posts mentioning the physical surroundings and linking to the past. \name posts have slightly more mentions in other aspects. (b) \name posts are slightly inclined to mention no more than two aspects compared to Forum. (c) \name has considerably more posts about food and schoolwork, and the forum has considerably more posts about recollection and information inquiry. The forum has slightly more posts in all other topics than \name{}.}
\end{figure}

Statistical data also supported the user patterns. The word counts of posts in the \name and the forum systems had averages of 7.2 and 22.5 and standard deviations of 7.39 and 22.46, respectively. Furthermore, our content analysis identified five aspects that emerged in memory posts, including specific events, surroundings, the author's emotions, opinions, or recollections of the past. A deeper look into their distributions also verified that posts in \name were more oriented towards the events or the writer's thoughts at the present, whereas forum posts were more oriented to what the surroundings had been like in the past (\autoref{fig:content:percent-aspect}). Also, fewer aspects were mentioned within one post in \name (\autoref{fig:content:n-aspect}). Lastly, we categorized each post into a topic (\autoref{fig:content:topic}). The distributions showed that the forum posts covered a more comprehensive topics, whereas the \name posts were mainly focused on schoolwork and food that participants were currently attending to.

As we were observing user activities during the field study, we noticed that the shorter and seemingly random posts in \name challenged our original expectation of ``collective memory'': they did capture the traits of the campus community to some extent, but we researchers were initially unsure about the extent of additional values they could add to the collective memory reservoir.
In response to this further thought during the interview, A8 refuted, ``\textit{When I wrote this [`handsome basketball' post], I did \emph{consciously} think that it was an important and worth-remembering part of our collective memory. It reflects the humor of us students here\dots{} The vibe, the atmosphere\dots{} I sensed it and I thought that is good for the community to remember.}''

In sum, such use patterns reflected two interesting points. First, the constituents of collective memory are not only the ``memorial'' narratives talking about the past, but also the ``memorizing'' acts documenting the presence. Second, although users have sought meanings and values for the collective from the memories they post, the meaning-making component may not always be present in the content. Especially when memorizing the ephemeral presence, users tend to write briefly with only the happenings, but not their interpretations.

\subsection{Participants' Comprehension of Memory Posts}\label{res:def}

Apart from writing, we have also uncovered a wealth of diverse opinions regarding participants' perceptions and meaning-making of each other's memory posts, and they were organized into three aspects.

The first aspect is the choice of topics. Some participants emphasized ``large events'' that a considerable number of community members have experienced together synchronously (A2, A9, A10, B4) or the historical transformations of the community (A4, A10).
A9 raised two examples, \textit{``An annual university-scale concert will directly influence all the campus members in that year and also indirectly strike resonances among the people involved in previous years. Also, imagine the sports societies representing the university attain great achievements. That will impact every campus member watching it.''}
Meanwhile, some also preferred everyday personal experiences that people share among the community either synchronously or asynchronously (A8, A10, B5).
B5 explicitly disfavored the \textit{``grand narratives like the university's anniversaries''}:
\textit{``I am more resonated by detailed personal connections like `we sat on the same chair in the canteen' and `we worked hard for the same deadline'.''}
A10, seeing both as valuable, thought \textit{``small things probably impact and connect a wider range of campus members, but in a less profound way.''}

The second aspect is how the individual memories were connected to the community.
Whether ordinary or extraordinary, participants generally accepted those memory posts about something the community had experienced together (A1, A2, A6, A7, A10, B1, B3) or the items or scenes that the community had in common knowledge (A6, B4, B5).
In addition, some remarked on the value of memories with interesting or unusual anecdotes, observations, and interpretations related to the community (A3, A6, A8, B4, B7).
Participants also stressed that content should be something the community could relate to and resonate with (A6, A7, B1, B7), instead of something too personal (A6, B3, B7).
In spite, there was an unclear boundary between ``personal'' and ``relatable'':
A7 recalled a previous campus exhibition of failure stories,
where she thought the stories were worth remembering, even though they were highly personal.
Her key argument was that the reflections from those stories were applicable to most community members, because people's lives are similar in a shared context, although the actual stories might be different and not relatable.
B1 added, \textit{``It made me think about the difference between `relatable' and `being touched or moved'. I meant `relating to' the exact same routines or experiences between community members. But there are also things that I only `sympathize' with emotionally. I think the former offers me deeper connections.''}

This led to the last aspect of the content's form.
How the memory stories were told could affect other people's perceived connection between them and the community.
This was mainly related to the memory posts about the present, as those about the past were usually meticulously written (see \autoref{res:trait}).
Some participants preferred posts with in-depth reflections and opinions (A3, A7, A9, B1, B3, B7), sometimes preferably long and elaborated (A7, B1).
A9 said, \textit{``The current (\name) content is not deep enough. I can only glimpse people's daily lives. I would expect more meaningful posts worth reading.''}
However, some preferred authentic self-expressions (B1, B2, B5, B6), which were not necessarily complex or extended (B2, B6).
B2 opined that \textit{``My (forum) platform appears as if people are talking to themselves, which makes me feel not part of them. The content on [\name{}] looks more `outgoing' and `alive', and I can personally connect with them.''}
B1 added that \textit{``Simple a memory of \textup{`Rushing to the bus! Two minutes left!'} is relatable to me. I can already recall a similar experience and sense the author's nerves.''}

\edit{From the questionnaire, we did not see significant differences in whether either group's content was considered more worth remembering (\(U=191.0, p=0.763\), effect size 0.543), relevant to the campus community (\(U=170.5, p=0.763\), effect size 0.482), or more contributive to the collective memory (\(U=154.0, p=0.426\), effect size 0.421).
  This lack of statistical differences reflects our qualitative findings: participants' diverse understandings of collective memory meant that no single content style or design resonated universally.
  It also implies that memories content about the past and the present are often expected to be presented in different narratives styles.
}

\subsection{Features and Usability}

\edit{
  In previous subsections, we have identified people's diverse expectations of the collective memory content.
  The lack of statistical significance in the Campus Connected Scale and the Collective Self-Esteem Scale (\autoref{tab:survey}) also supports the finding that neither design was universally superior at fostering community bonds.
Thus,} this subsection revisits the features in both systems, reflecting on their different affordances and pragmatic concerns in supporting collective memory co-contribution in any style.

With people's diverse expectations of the collective memory content, this subsection revisits the features in both systems, reflecting on their different affordances and pragmatic concerns in supporting collective memory co-contribution in any style.


%

\subsubsection{Content Organization}

The most prominent difference between the two systems is the way content is organized (location-based versus location-agnostic).
In the interview, participants confirmed and added that the location was important information about a memory (A3, A4, B1).
However, more ambivalent attitudes emerged when it came to the locative interaction mode. Although some acknowledged that it could evoke more memories about the community's shared context (A10, B1) and encourage more writing (A3, A7, A9, A10) and reading (B4, B5) in daily lives, participants were not always willing to dedicate physical efforts to finding posts in their daily routine, thus seeing fewer posts that are not around their commute routes (A2, A4, A9).
The unwillingness to explore had multiple reasons, including personal arrangements (A4, A9), familiarity with the campus (A3), and unattractiveness of the posts (especially at the most commonly used places: canteens and workspaces, see \autoref{res:trait}) (A2).
\edit{This is similarly reflected in the questionnaire. The location-based content organization is rated significantly more logical (\(U=70.0, p=0.001\), effect size 0.207), navigable (\(U=75.0, p=0.001\), effect size 0.213), and linked to the real-world campus environment (\(U=110.5, p=0.035\), effect size 0.282); but in the meantime, we could not identify significant differences in people's impressions (\(U=194.0, p=0.696\), effect size 0.532) and motivation (\(U=133.5, p=0.156\), effect size 0.344).}
Concerning the transactive nature of collective memory~\cite{wegnerTransactiveMemoryContemporary1987}, it highlights the different affordances of the two designs.
On the one hand, the locative foregrounds the location-situatedness of the posts (A1, A5, A4, B3) and \textit{```keeps [things] fresh' for a longer time''} (A4). It may be more engaging if the experience is more tied to the synergy of the content and the physical exertion.
On the other hand, the forum design readily delivers the most recent information and provides a more efficient experience (A7, A9, B5, B7). It is more capable of hosting standalone, content-focused narratives.

\subsubsection{UI Style and Interactivity}

To some participants, the postcard design in \name strengthened the theme of communicating memories (A3, A10), and it sometimes intensified emotions when reading others' posts (A6, A10).
B3 and B7 also expressed more interest in writing memories on \name because it reminded them of stamp collections and sticky note walls.
The postcard's color-changing effect also strengthened an impression of memory (A2, A6) and time (A3), creating a sense of mystery out of the less illegible photos.
However, since many memories people posted were close to the present (\autoref{res:trait}), A1 also noted that there might not always be a sepia filter to represent the distant past. B3 was also concerned about whether it would be too visually monotonous.

Next, the tilt-to-open interaction added engagement with the memory (A2, A9) as if manipulating physical memorial objects (A7, A8).
Some participants even felt a sense of ritual and treasuring beyond simply reading (A8, A9).
Nevertheless, sometimes it interrupted participants' flow of continuously browsing content (A3, A4, A5). The gyroscope was also inherently more inaccurate if the mobile phone was held near-vertical. Hence, it often failed to detect further tilting-up actions if users started with a near-vertical position (A1, A3, A5, A6, A7).

\begin{figure}[ht]
  \centering
  \begin{subcaptionblock}[c]{0.2\textwidth}
    \includegraphics[width=\textwidth]{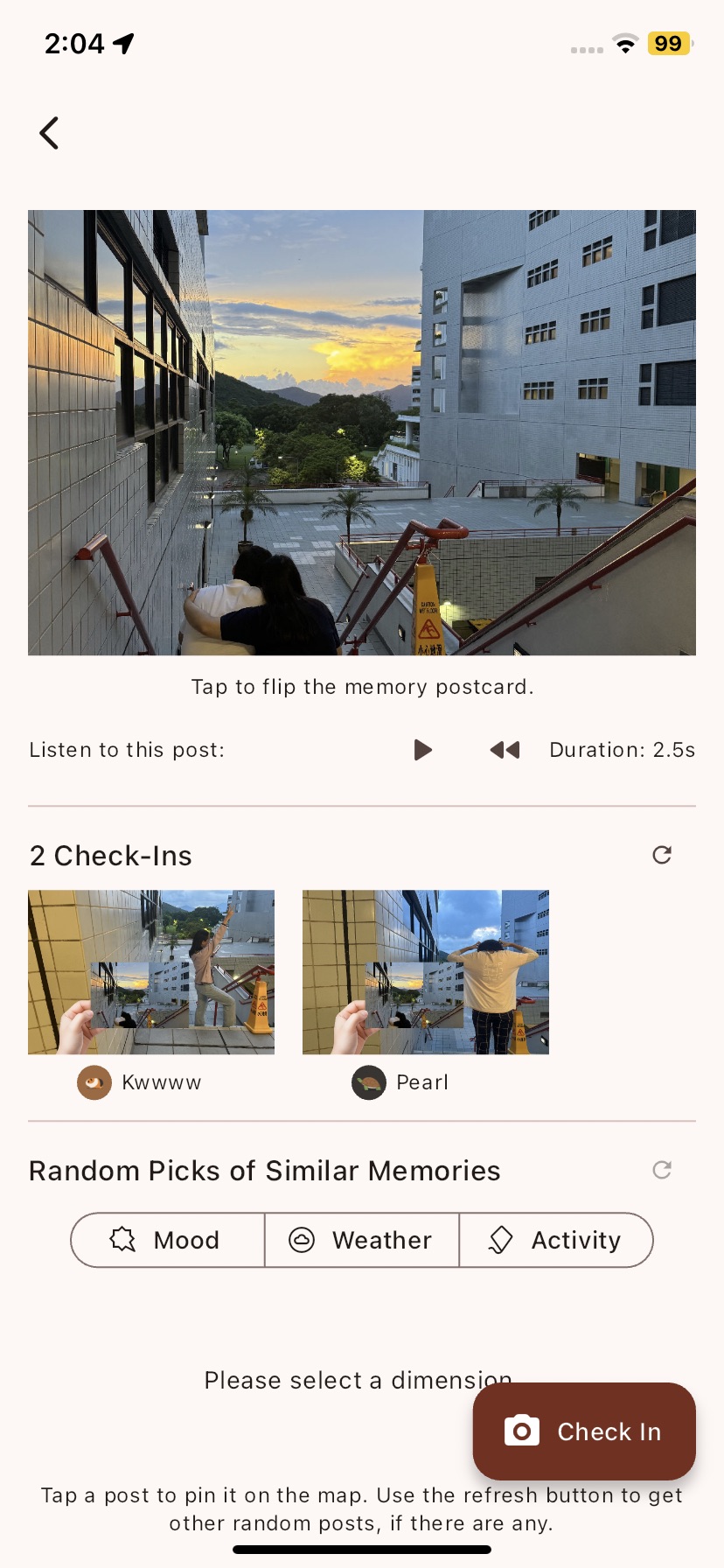}
  \end{subcaptionblock}
  \hspace*{0.4cm}
  \begin{subcaptionblock}[c]{0.6\textwidth}
    \includegraphics[width=\textwidth]{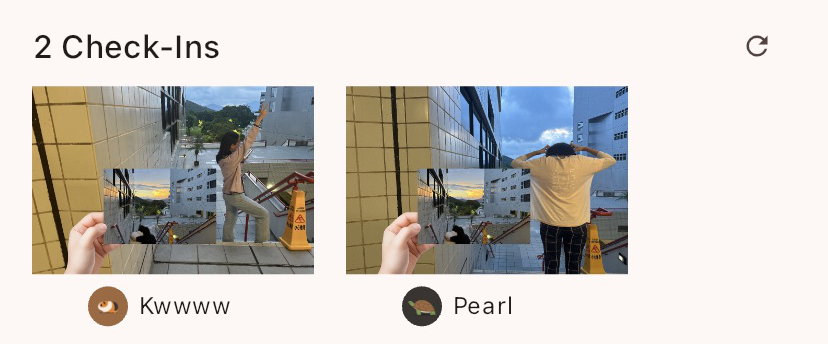}
  \end{subcaptionblock}
  \caption{A playful and creative check-in. Two participants ``teased'' the couple in the postcard photo.}
  \Description{A memory postcard about the sunset and the backs of a cuddling couple. The two check-in photographs are of users making funny, teasing body gestures at the couples in the postcard.}
  \label{fig:playful-check-in}
\end{figure}

The check-in photography was seldom used during the field study compared to the forum's ``like'' button. Although it could convey the message of ``I have been here'' (A8), it required too much extra physical effort (A2), thus did not always match one-click likes (A7).
B1 imagined that users could play creatively with this feature, like eating a postcard.
In fact, when we continued to serve \name after the field study, such playful interactions did emerge (\autoref{fig:playful-check-in}).

Thus, a takeaway from the three concrete skeuomorphic features is \edit{the trade-off between engagement and usability.
  On the one hand, \name interactions did create occasional friction. This is mainly reported qualitatively, especially when applied to short and present-oriented narratives in collective memory. But quantitatively, the two systems remained at similar levels of pragmatic qualities, with no significant differences in efficiency (\(U=154.0, p=0.404\), effect size 0.423), ease of use (\(U=224.0, p=0.158\), effect size 0.626), clarity (\(U=190.5, p=0.767\), effect size 0.547), etc.
One the other hand, the skeuomorphic design of \name was perceived as significantly more interesting (\(U=118.0,p=0.049\), effect size 0.317), inventive (\(U=75.5,p=0.001\), effect size 0.185), and cutting-edge (\(U=285.0,p=0.003\), effect size 0.225) than the forum.}

\subsubsection{Tagging and Memory Recommendation}

Several participants found the tag suggestions accurate and useful (A1, A7, A9, A5), and B5 thought \name promoted better-rounded information than the forum system one through specific suggestions on this informative metadata.
B3 also added that, from a reader's perspective, presenting tags in visuals triggered similar impressions to verbal descriptions in the main content.
The specific tag dimensions also brought about an effective heuristic when participants had limited time to discover posts (A3, A7, A9).
A3 recalled that \textit{``Once I had just had McDonald's at midnight, and I happened to be recommended another post about McDonald's last night. At that moment, I felt a sudden surprise and sympathy for that user.''}
It was also suggested that the system could encourage serendipitous discovery by recommending memories of diverse topics (A4) and new memories (B1).

\subsubsection{Other Features}\label{res:usability:other}

Among the \name users, the majority seldom used the audio-related feature during the study.
Since their system usages mostly happened along their commute routes, sending and playing audio recordings could be awkward when people were around (A2, A6, A7), and some participants were more used to sending text in such ad-hoc situations (A2, A6, A7, A10).
The audio-related features were originally designed to afford authoring and consuming long memory posts. However, concerning the more dominant impact of content organization methods on the memory post styles, the audio-related features became marginal.
Other than allowing memory documentation in audio modality, a viable alternative can be recording ambient sounds in conjunction to enhance the reading experience (A7, B5).

Additionally, participants have suggested several overview features to reinforce the perception of collective memory,
including presenting daily statistics of users' memory posts (A8, A9),
summaries of each region's memories and vicissitude (A2, A4),
and popular regions (A10).
Some also demanded more instant social feedback of their input (A7, A9, B5) and dedicated discussion spaces for specific topics that emerged in the long run (A7, A10, B2, B3, B7)

\section{Discussion}

\edit{
  With our unique bottom-up perspective, we have obtained an insightful, user-centered understanding of grassroots collective memory conceptualization and co-contribution practices surrounding collective memory.
  Their diverse expectations of topic choices, narrative styles, temporal situations, and values to the community have greatly addressed \ref{rq:conceptualization} and \ref{rq:practice}.
  In this section, we further extract design considerations and answer \ref{rq:considerations}, for future researchers to transform these conceptual insights into practical advice.
}

\subsection{Embracing Diverse Interpretations of Collective Memory}\label{dis:def}


Among all the manifold conceptualizations of collection memory, our study especially discovers one debate about the constituents of collective memory that is less focused in previous literature: some participants advocate \textbf{the retrospective contemplation and emotional expression (which benefit the group with relatable experience and monumental events)}, whereas some participants emphasize \textbf{the brief and spontaneous documentation of the current moment as a history for the future (which complete the knowledge of the group's vibe and easy-to-miss details)}.
This discovery also demonstrates how the HCI perspective and a ``bottom-up'' approach (\autoref{sec:rw:cm}) adds value to sociological studies of collective memory because it originates from the procedure of communal co-contribution and day-to-day interpersonal interaction.
Meanwhile, some previous sociological debates are also present in our findings: one is regarding the community's monumental or symbolic highlights~\cite{schwartzAbrahamLincolnForge2003} versus the minutiae of everyday lives~\cite{halbwachsSpaceCollectiveMemory1950}; the other is regarding common knowledge and co-experiences~\cite{schudsonDynamicsDistortionCollective1995} versus relatable individual experiences and thoughts~\cite{zerubavelSocialMindscapesInvitation1997}.

These conceptual discrepancies do lead to different design focuses.
For example, for long and expressive memory stories of a historical event or a personal experience, more poetic and metaphorical interactions like the box-opening effect can stimulate recollection and empathy, encouraging serious engagement. Categorization and association of memories based on personal contexts (e.g., the dimensions of emotions, activities, and weather) are also useful to poetically connect similar minds together. However, more direct access, such as a traditional forum feed, will allow sufficient social feedback to acknowledge authors' efforts and link similar people together.
Meanwhile, for quick, ad-hoc memory records, collective memory may surface from the swarm of similar memories. Thus, the system can support efficient input and auxiliary display of extra contextual information. Locative interaction also shines because the process itself is astounding when one physically discovers and witnesses swarming content at a special spot.

Eventually, these differences entail a design consideration: to \textbf{functionally support different interpretations and practices, allowing their co-existence within a single community.}
Through feature modularization and adaptation to content styles, future designs can strive to showcase the meanings and maximize the impressions of different memory styles, reflecting the vibrancy of the community.

\subsection{Interpretative Diversity as a Reflective Design Opportunity}\label{dis:reflect}

Following our reflection above, we further propose that designers can readily pass the embrace of interpretative diversity on to users.
That is, future designs can nudge users to make meaning not only from their preferred styles of narrations and interactions but also from more utterances in the prolific community intelligence.
This proposal follows the spirit of \emph{reflective design}~\cite{sengersReflectiveDesign2005}: the ultimate goal is not only about researchers' own reflection on the outcome, but also to \textbf{continuously bring this concept to users' conscious awareness and kindle their critical reflection on the constituents of collective memory and their impact}.

During our field study, we indeed observed the fluidity in some participants' definitions of collective memory (A7, B7)---resulting from their contemplation of previously less preferred content types throughout their usage.
For example, A7 started to realize that some personal stories in an autobiographical style also inspired the community with wit and new perspectives, and they were also integral portions of the collective memory.
B7 started to acknowledge the value of the shorter, less informative posts, as the process of ``joining the trend'' and ``jotting'' might strengthen people's shared memory.
Hence, explicitly nudging users' reflection may also influence how they remember the group.
Even more, the shaping and evolution of collective memory's definition landscape could be part of the community's collective memory as well.

\subsection{Tensions and Potential Risks of the Design}
Collective memory inherently involves the community's social interactions of memory exchange~\cite{wegnerTransactiveMemoryContemporary1987}. At a bigger picture, this process is ultimately for the communal good, blending narratives together to form a general impression instead of spotlighting specific individual perspectives~\cite{heuxCollectiveMemoryAutobiographical2023,halbwachsSpaceCollectiveMemory1950,harrisCollaborativeRecallCollective2008}.
However, it is not always the same during individual users' active co-contribution process.
This subsection discusses some potential consequences and risks when designing for collective memory co-contribution.
Either hinted by our study participants or drawn from literature, they delineate some important design considerations or interesting research agenda for future HCI scholars.

\subsubsection{Tension between the Communal Cause and the Social Concerns}\label{sec:consideration:social}

During the interview, we have observed some participants' social concerns when engaging in the system (A2, A7): a worry of being judged when a strong social attribute of the platform is perceived~\cite{huangSharingFrissonsOnline2024}.
Such a concern is more salient if the user's own interpretation of collective memory involves authentic and detailed sharing of personal experiences and feelings.
In the scope of collective memory, we note that this tension goes beyond a mere feature choice. It reflects \textbf{the blurry line between individual content sharing and collective sensemaking, as well as how the social concern may challenge the information gathering stage}. As previous ``top-down'' perspectives often focus on the transactive flow of memories and the formulated collective memory outcome~\cite{liuGrassrootsHeritageCrisis2010,wegnerTransactiveMemoryContemporary1987,halbwachsSpaceCollectiveMemory1950,lakCollectiveMemoryUrban2019}, our grassroots findings complement the HCI bigger picture with the stage before memories are exchanged and collectively processed.
In this regard, we also call for more future design explorations to lower the mental barrier---for example, making use of gamification~\cite{liuCoArgueFosteringLurkers2023}, accentuating the aggregated summarization instead of individuals' utterances~\cite{huangSharingFrissonsOnline2024,robinsonChatHasNo2022}, channeling more social acknowledgement and reciprocity~\cite{caraban23WaysNudge2019}.


\edit{
  \subsubsection{Potential Privacy Concerns of Geolocation Data}
  Apart from the social concerns, the richer feature set in \name also leads to more possibilities of privacy concerns. For example, the geo-tagging of personal memories and photos may inherently encode sensitive information, such as identifiers, location and activity histories, and absence information, potentially resulting in unconscious information leaks~\cite{ruizvicenteLocationRelatedPrivacyGeoSocial2011,freniPreservingLocationAbsence2010}.
  Such concerns are not present in our interview even though we asked the participants explicitly, possibly because we have addressed them during the onboarding briefing.
  Still, future real-world deployments are suggested to clearly state the potential risks of involving sensitive geolocation data, ensuring users' privacy awareness~\cite{barthPrivacyParadoxInvestigating2017}.
  In addition, these deployments can also provide granular control over the information visibility and anonymity, satisfying varying privacy preferences of different users in different communities.
  For example, the system can offer to configure the users to whom a memory post is visible, mask faces in photographs when viewed by different people, and replace original voices with AI-generated ones.

  \subsubsection{Potential Negative Social Phenomena}
  In our study, we have set a clear rule that toxic, hostile, and other negative social acts in both systems are strictly prohibited.
  However, in real-world deployments, social factors are inherently attributed to such collective memory facilitation systems (as also addressed in \autoref{sec:consideration:social}), and negative social acts may inevitably occur, either consciously or unconsciously.
  For example, people may have conflicting perspectives and memories regarding the same event~\cite{houYouDontUnderstand2025,momennejadBridgeTiesBind2019}, and one person's memory may be another person's trauma~\cite{zhangCrisisCollectiveMemory2020}.
  Besides, within an enlarged user base, in-groups and out-groups may form, resulting in memory content barely understood outside subgroups~\cite{maattaEverydayDiscourseSpace2021}, or an increased toxicity level~\cite{parkYouChangeWay2024}.
  Thus, if a community platform is dedicated to collective memory development, it is suggested to actively detect potential harmful and toxic content before publishing, and prompt the user to refine it.
}

\subsection{Locative Interactions for Community Members' Devoted Narration}

We also revisited our design of \name and our initial expectation of delivering locative narrative experiences of long, expressive, and storytelling UGC (which we name ``devoted narrations'' below).
Despite our design, participants were still sometimes reluctant to put effort into such devoted narrations.
Reflecting on the integration of the immersion and engagement of locative interactions into devoted narrations---especially on UGC platforms for communal empathy and meaning-making---we specifically highlight two key design considerations: \textbf{leveraging the physical affordances in the environment} and \textbf{encouraging dedicated usage sessions}.


The first point is based on the significantly more user activities at canteens and workspaces because of participants' ``resting'' state at those locations. Therefore, such physical affordances of different locations can be an effective element to utilize in future locative systems.
It is similar to Millard et al.'s utilization of the terrain's geological traits~\cite{millardCanyonsDeltasPlains2013}---except that they focused on content consumption experiences and did not consider the functional affordances of geolocations.
In our case, for example, designers can specially enhance the ``resting'' locations with extra motivation and authoring preparation---as if checkpoints in games.

Secondly, considering the physical efforts of \emph{initiating} one locative narration session, authoring several posts in a dedicated session may be more preferable than fragmented writing in everyday routine.
Dedicated authoring sessions can be functionally supported by the ability to create more fluent experiences, such as ``memory trajectories''. They can serve as both a logical thread connecting narrative pieces and the building block for immersive exploratory experiences~\cite{cheverstSupportingConsumptionCoauthoring2017,lovlieYouAreOne2012,benfordInteractionTrajectoriesDesigning2009}.
Furthermore, designers can build up users' intrinsic motivation through reward systems like gamification or amplified social feedback~\cite{liuCoArgueFosteringLurkers2023}, as well as an enhanced recommendation system based on users' narration activities and writing styles.

Several existing HCI works have shown the power of locative UGC platforms for the community's cause, as the content readily augments the geographical reality and takes part in placemaking~\cite{cranshawJourneysNotesDesigning2016,cheverstSupportingConsumptionCoauthoring2017}. But many either focused on short and ad-hoc content~\cite{juBreadcrumbSNSAsynchronous2015,bowserPrototypingPLACEScalable2013,mcgookinStudyingDigitalGraffiti2014,changTourgetherExploringTourists2020} or did not evaluate the experience of serious content authoring~\cite{cranshawJourneysNotesDesigning2016,cheverstSupportingConsumptionCoauthoring2017}.
Therefore, our design process and findings also generally add insights to the mobile HCI community on effectively supporting locative and devoted narrations.

\edit{
  \subsection{Generalizability}
  In this subsection, we reflect on the ways our findings can be generalized to a broader context.
  Methodologically, our dual-system setup strategy incorporates both a well-established, content-oriented design and an inchoate, interactive one.
  Our research questions also consider how different designs may affect users' behavioral and interpretational patterns.
  As such, we have striven to ensure that our findings are applicable to a wider range of platforms and designs.

  Concerning the findings, this work aims to conceptualize concrete memory outputs from the community and probe into people's rationale behind the collective memory construct itself.
  Thus, the generalizability of our findings lies in the fundamental dimensions of diversity (e.g., present vs. past, monumental vs. everyday) that any system aiming to support collective memory must accommodate from the outset.

  Finally, the bottom-up perspective itself demonstrates great values in uncovering grassroots understanding of this community-level concept and their self-positioning therein.
  We hypothesize that similar bottom-up approaches can be applied in other HCI studies that aim to bridge sociological analyses and user-centered design considerations.
}

\subsection{Limitations}\label{dis:lim}

\edit{
  First of all, it is crucial to highlight our study's deliberately bottom-up approach.
  Since this approach is novel and lacks mature technological support, our investigation is exploratory in nature.
  The contributions of this work are centered on generating new observations and design insights for an under-explored area, rather than producing concrete solution systems or comprehensive design frameworks.
}

\edit{
  Second, we acknowledge that the two-week duration may not fully capture the community's dynamics (such as entering, leaving, retention, and returning).
  It may neither capture the potential variation in usage rates as the novelty effect may gradually diminish.
  Nevertheless, this paper primarily focuses on exploring the grassroots conceptualizations and practices, instead of proving the systems' capability to support the complete formation of collective memory.
  Thus, the two-week duration has already helped us investigate the crucial bootstrapping phase of a community's conscious memory co-contribution process.
  And we can contribute an in-depth analysis of the initial conceptualizations, negotiations, and practices that occur when a community joins this process.
  That being said, we still acknowledge the significance of longer-duration studies in completing the domain understanding. We plan to conduct such an extensive, longitudinal study as the next step of this research.
}

\edit{Third, the participant pool is still small compared to a real-world, campus-scale community,} and participants' demographics are skewed toward research postgraduates.
The limited coverage may affect the scope of the collective memory content and behavioral patterns that we can capture. The constitution of the user group may also subtly affect their communicational behaviors on the platforms.

Besides, people's commenting behaviors were not analyzed, as we were more focused on the creation and consumption of new memory posts that participants contributed. However, the frequency and content of comments can also reflect users' subsequent digestion of and reaction to the collective memory. In future work, we are also interested in observing users' commenting behaviors as well as exploring alternative designs of the comment feature to better facilitate the co-contribution of collective memory.

\edit{Finally, as we primarily focused on the bottom-up probing and exploration of users' conceptualizations and practices, we did not conduct a quantitative and comprehensive comparison between the user behaviors in our systems and other existing collective memory-related systems.
Although we have observed that the content styles in our forum system are similar to those in other studies about collective memory development in general-purpose social platforms (\autoref{res:trait}), a more systematic comparison may better ground our findings and visualize how they add to the current domain knowledge.}

\section{Conclusion}

In this paper, we explore grassroots understandings and practices of collective memory co-contribution in a university setting. And we approach it from a ``bottom-up'' perspective, intended to differ from the more common ``top-down'' one that focuses on its holistic relation to the community. This perspective \edit{is uniquely suited} to gather more human-centered insights about community members' conceptual and behavioral patterns concerning their collective memory, its constituents, and its co-contribution process.
To this end, we conducted an exploratory, two-week, mixed-methods field study with 38 participants.
With two design probes---one in the form of a traditional online forum and the other with locative, and skeuomorphic design condensed from existing literature---we inquired into people's conceptualization of collective memory and its association with the choice of the form.

\edit{The primary contribution of our exploratory investigation is the meaningful identification of} participants' diverse interpretations of collective memory and the forms of its constituents.
Our findings are mainly in the dimensions of content styles (extensive expression versus brief utterance) and temporal positionings (writing past memories for the present versus writing the present as a memory for the future), and they also echoed previous debates on choices of the topic (monumental events versus everyday minutiae) and different links to the community (common knowledge versus relatable thoughts).
The different conceptualizations subsequently affected users' expectations of the memory co-construction system and the insights they would obtain from memory exchange and collective sensemaking.
As final and practical design considerations, we suggest that future HCI research strive to embrace the interpretative and interactional diversity in supporting systems, encouraging users' critical reflections of such diversity as part of the experience, and lowering users' social concerns to motivate collaboration. With reflections on our design, we also offer practical considerations for future HCI systems that aim to foster collective memory development and community-driven memory exchange.

\begin{acks}
  This project is partially supported by the Hong Kong SAR Research Grants Council's Theme-based Research Grant Scheme (Project No. T43-518/24-N).
\end{acks}

\bibliographystyle{ACM-Reference-Format}
\bibliography{main}


\end{document}